\documentclass[a4paper,12pt]{article}

\usepackage [T1]{fontenc}
\usepackage {calligra}
\usepackage [T1]{pbsi}
\usepackage{CJKutf8} 
\usepackage[colorlinks=true,linkcolor=blue]{hyperref}
\usepackage{amsmath,amssymb,amsfonts} 
\usepackage{graphicx}                 
\usepackage{color}                    
\usepackage{hyperref}                 
\usepackage[left=1cm,top=1cm,bottom=1cm,right=1cm,nohead,nofoot]{geometry} 
\definecolor{red}{rgb}{1,0,0}           
\definecolor{green}{rgb}{0,1,0}
\definecolor{blue}{rgb}{0,0,1}
\definecolor{darkblue}{rgb}{0,0,0.5}
\definecolor{lightblue}{rgb}{.5,.5,1}
\definecolor{lightgray}{gray}{.87}          
\definecolor{Dark}{gray}{.20}
\definecolor{pink}{rgb}{.95,0.82,0.92}  
\definecolor{yellow}{rgb}{1,1,0}
\definecolor{lightyellow}{rgb}{1,1,.5}
\definecolor{purple}{rgb}{0.7,0,0.85}
\definecolor{darkgreen}{rgb}{0,0.5,0}
\definecolor{orange}{rgb}{0.8,0.2,0.2}
\def \be {\begin{equation}}
\def \ee {\end{equation}}
\def \bea {\begin{eqnarray}}
\def \eea {\end{eqnarray}}
\def \nn {\nonumber}

\def \rr {\raise.35ex\hbox{\small $\prime$}\kern-.17em{\mbox{\large $\imath$}}}
\def \del {\partial}
\def \dels {\partial\kern-.5em / \kern.5em}
\def \As {{A\kern-.5em / \kern.5em}}
\def \Ds {D\kern-.7em / \kern.5em}

\def \a {\alpha}

\def \b {\beta}

\def \g {\gamma}

\def \d {\delta}
\def \eps {\epsilon}
\def \m {\mu}
\def \n {\nu}

\def \lam {\lambda}
\def \Lam {\Lambda}
\def \s {\sigma}
\def \r {\rho}

\def \th {\theta}

\newcommand{\ba}{\begin{eqnarray}}
\newcommand{\ea}{\end{eqnarray}}

\newcommand{\ur}{\underline{\rho}}
\newcommand{\ttg}{\mathtt{g}}

\newcommand{\tA}{\tilde{A}}
\newcommand{\ta}{\tilde{a}}
\newcommand{\tB}{\tilde{B}}
\newcommand{\tF}{\tilde{F}}
\newcommand{\tZ}{\tilde{\mathcal{Z}}}
\newcommand{\tL}{\tilde{\Lambda}}

\newcommand{\bF}{\mathbf{F}}

\newcommand{\tchi}{{\tilde \chi}}
\newcommand{\tphi}{{\tilde \phi}}
\newcommand{\tY}{{\tilde Y}}
\newcommand{\sD}{{D\!\!\!\!\slash}}

\def \newchi {\chi}
\def \newphi {\phi}
\def \newY {{Y}}

\def \z {(0)}
\def \KK {(KK)}

\newcommand{\Tr}{\mathrm{Tr}}

\newcommand{\solution}[1]{}

\newcommand{\remove}[1]{}

\newcommand{\CK}{^{\tiny [CK]}}
\newcommand{\lCK}{_{\tiny [CK]}}

\begin{document}

\pagestyle{plain}

\begin{CJK}{UTF8}{bsmi}

\begin{titlepage}

\begin{center}

\begin{flushright}
UT-14-39
\end{flushright}

\vskip 12mm

\textbf{\LARGE 
Aspects of Effective Theory \\
\vskip.0cm
for Multiple M5-Branes \\ 
\vskip.3cm
Compactified On Circle
}

\vskip 2cm
{\large
Pei-Ming Ho$^\dagger$\footnote{
e-mail address: pmho@phys.ntu.edu.tw} and 
Yutaka Matsuo$^\ddagger$\footnote{
e-mail address:
matsuo@phys.s.u-tokyo.ac.jp}}\\
\vskip 2cm
{\it\large
$^\dagger$
Department of Physics and Center for Theoretical Sciences, \\
Center for Advanced Study in Theoretical Sciences, \\
National Center for Theoretical Sciences, \\
National Taiwan University, Taipei 10617, Taiwan,
R.O.C.}\\
\vskip 3mm
{\it\large
$^\ddagger$
Department of Physics, Faculty of Science, University of Tokyo,\\
Hongo 7-3-1, Bunkyo-ku, Tokyo 113-0033, Japan\\
\noindent{ \smallskip }\\
}
\vspace{60pt}
\end{center}
\begin{abstract}

A supersymmetric non-Abelian self-dual gauge theory 
with the explicit introduction of Kaluza-Klein modes 
is proposed to 
give a classical description of
multiple M5-branes on $\mathbb{R}^5\times S^1$.
The gauge symmetry is parametrized by Lie-algebra valued 1-forms 
with the redundancy of a 0-form,
and the supersymmetry transformations without gauge-fixing are given.
We study 
BPS configurations involving KK modes,
including M-waves and 
M2-branes with non-trivial distributions around the circle.
%
Finally, 
this supersymmetric gauge theory of two-forms 
can be equipped with more general non-Abelian gerbes in five dimensions.

\end{abstract}

\end{titlepage}

\setcounter{footnote}{0}

\section{Introduction}
\label{intro}

In the past few years, 
there has been growing interest in 
finding a low energy effective theory for 
multiple M5-branes in M theory
\cite{r:Bekaert1}--\cite{Palmer:2014jma}.
One of the many approaches to the problem is to consider
M5-branes compactified on a circle of finite radius $R$.
(In the end,
you can take the decompactification limit $R\rightarrow \infty$
for the uncompactified theory.)
Via double dimensional reduction,
when $R \rightarrow 0$, 
M5-branes become D4-branes,
which are described by the 5-dimensional super Yang-Mills theory.
This duality serves as an important constraint on 
the model for multiple M5-branes.

At the same time,
a model of multiple M5-branes should admit the configurations
in which all M5-branes are well separated from each other
so that they are all decoupled.
In this limit, 
the model should be described by 
multiple copies of the single M5-brane effective theory,
which has a well-known Lagrangian \cite{r:PS,r:PST} 
with an Abelian gauge symmetry
(with or without compactification).
This is another important constraint on 
the multiple M5-brane theory.

In addition, 
the world-volume theory of M5-branes is expected to have
the $(2, 0)$-superconformal symmetry in 6 dimensions.
Although it is possible that only part of the supersymmetry is manifest
in a Lagrangian formulation \cite{r:SSW,r:Chu,r:SaemannWolf,r:Bonetti},
the same field content (more precisely the dynamical degrees of freedom) 
should agree with that of the $(2, 0)$-theory.

In our opinion, the most important feature of M5-branes 
is the gauge symmetry of a 2-form gauge potential.
While the Abelian theory for such a gauge symmetry is well understood
both in physics and mathematics \cite{r:AbelianGerbe},
the non-Abelian counterpart is rather mysterious.
In mathematics, 
there is still no consensus about the precise definition 
of non-Abelian gerbes \cite{r:BM,r:BH}.
In physics, 
the construction of a satisfactory theory for non-Abelian 2-form gauge potential
is usually obstructed by various no-go theorems
\cite{Teitelboim:1985ya,Henneaux:1997ha,r:Bekaert1,r:Bekaert2,r:Bekaert3,r:CHT,r:MM5b}.
\footnote{
See also \cite{Gukov,Kapustin} as a different class of applications
of 2-form gauge theory in physics so that 
the no-go theorems are not relevant.
}
A crucial difference between the ordinary gauge symmetry for 1-form potentials 
and that for 2-form potentials is that 
the latter has a redundancy in the gauge transformation laws.
How to non-Abelianize the gauge symmetry without losing 
this ``gauge symmetry of gauge symmetry'' is the key issue of the problem
in order to have the correct number of degrees of freedom.
This is perhaps directly connected to the core of the mysteries about M5-branes,
which offer an opportunity to guide us to 
significantly expand our understanding of the notion of gauge symmetry.

In fact, 
there is already a non-Abelian gauge theory 
for a 2-form potential.
It is the effective theory for a single M5-brane 
in the background of large $C$-field \cite{C-field}.
The non-Abelian algebra is characterized by the Nambu-Poisson bracket 
as a result of the $C$-field background.

In previous works \cite{r:HHM,r:Kuo-Wei,r:HoMatsuo},
a model was proposed for the gauge field degrees of freedom 
in a system of multiple M5-branes.
Its 6-dimensional base space is compactified on a circle of finite radius $R$,
and it satisfies the following criteria:
\begin{enumerate}
\item
When KK modes are removed on dimensional reduction,
it reduces to the Yang-Mills theory, 
the gauge field sector of multiple D4-branes.
\item
When the gauge group $U(N)$ is replaced by $U(1)^N$,
it reduces to decoupled multiple copies of 
the 6D self-dual gauge theory.
\item
It has a consistent non-Abelian gauge symmetry algebra 
\footnote{
The gauge transformation should be parameterized by a 1-form 
in the adjoint representation of the gauge group,
with a redundancy parametrized by a 0-form.
}
for a self-dual 2-form potential in 6 dimensions,
without any excessive physical degrees of freedom 
(such as an extra 1-form potential).
\end{enumerate}
This proposal \cite{r:HHM,r:Kuo-Wei,r:HoMatsuo} 
stands out as the only existing model
that has been shown to satisfy all three criteria above.
However, it misses the ingredients of matter fields and supersymmetry.
One of the purposes of this paper is to 
show that an existing proposal \cite{r:Bonetti} 
of the supersymmetric theory for multiple M5-branes 
\footnote{
The theory of \cite{r:Bonetti}
is derived from the framework of supersymmetric theories developed in
a series of works \cite{r:Bergshoeff}.
}
can be viewed as the gauge-fixed version 
of the supersymmetrization of this non-Abelian self-dual gauge theory.
It has the right field content, 
although only part of the $(2, 0)$-supersymmetry is manifest.
With SUSY, 
one may proceed to study various aspects of the system in more detail, 
such as supersymmetric classical configurations,
which are the other focus of the paper.

The plan of the paper is as follows.
We review and elaborate on the non-Abelian 2-form gauge theory 
\cite{r:HHM,r:Kuo-Wei,r:HoMatsuo} in Sec. \ref{Non-Abelian}.
We show that there are infinitely many conserved charges 
associated with the translation symmetry in the compactified direction.
In Sec. \ref{SUSY} we extend the supersymmetry algebra
proposed in Ref.\,\cite{r:Bonetti} to a larger algebra
that closes on the gauge transformation,
so that the former can be viewed as
the gauge-fixed version of the latter.
In Sec. \ref{Soliton},
we construct BPS configurations which involve KK modes,
including those describing 
M2-branes lying along a non-compactified direction
with non-trivial distribution in the compactified direction.
We take the large $R$ limit of these BPS solutions
and evaluate their behavior.
%
We also find BPS states corresponding to M-waves,
that is, propagating waves in the compactified direction.
In Sec. \ref{s:gerbe}, 
we point out that supersymmetric gauge theories can be defined
for more general set-up of non-Abelian gerbes in 4+1 dimensions
\cite{r:HoMatsuo}.
Finally,  in Sec. \ref{Comment},
we comment on other approaches to multiple M5-branes and conclude.

\section{Non-Abelian 2-Form Gauge Theory}
\label{Non-Abelian}

In this section, 
we review the gauge symmetry and action 
for the non-Abelian self-dual gauge field 
proposed in Refs. \cite{r:HHM,r:Kuo-Wei,r:HoMatsuo}.
We will also analyze the theory in more detail,
giving expressions for the Hamiltonian,
Poisson brackets and conserved charges.
An interesting feature of the theory is that 
there are infinitely many conserved charges, 
as a result of the property that 
all KK modes interact only through zero modes.

\subsection{The Non-Abelian Gauge Symmetry}

The base space of the theory is $\mathbb{R}^5\times S^1$.
The coordinates $x^\mu$ $(\mu = 0, 1, 2, 3, 4)$ are used for $\mathbb{R}^5$
and $x^5$ for $S^1$.
Naturally, 
the 2-form potential $B_{MN}$ 
($M, N = 0, 1, 2, 3, 4, 5$) is decomposed into 
two sets of components $B_{\mu 5}$ and $B_{\mu\nu}$.
It is also natural to decompose all fields into 
zero modes and Kaluza-Klein (KK) modes as
\footnote{
In fact, 
a quantity only has to be periodic up to 
a gauge transformation, 
so the decomposition of a field 
into zero modes and KK modes 
as in (\ref{decomp}) is not always possible.
We will comment on twisted boundary conditions
in Sec. \ref{comments}.
}
\be
\Phi = \Phi^{\z} + \Phi^{\KK}.
\label{decomp}
\ee

The gauge potential $B_{MN}$ and gauge transformation parameter $\Lam_{M}$
take values in a non-Abelian Lie algebra.
It should be $u(N)$ for $N$ M5-branes in flat spacetime.
We identify the Wilson loop (zero mode)
of the one-form gauge parameter $\Lam_M$
as a 0-form gauge parameter $\lambda$
\be
\lam \equiv 2\pi R\Lam_5^{\z} = \oint dx^5 \; \Lam_5,
\ee
which is independent of $x^5$.
With respect to this 5D gauge parameter $\lambda$,
we shall treat the zero mode of $B_{\mu 5}$ as 
the corresponding 5D 1-form potential $A_\mu$.

The non-Abelian gauge transformation law for the 2-form potential 
is defined by \cite{r:HHM}
\footnote{
$F_{\mu\nu}$ here is different from the $F_{\mu\nu}$ in Ref.\,\cite{r:HHM}
by an overall factor of $2\pi R$.
}
\bea
\d B_{\mu 5} &=& 
[D_\mu, \Lam_5] - \del_5 \Lam_\mu + [B_{\mu 5}^{\KK}, \lam],
\label{transf-Bi5}
\\
\d B_{\mu\nu} &=&
[D_\mu, \Lam_\nu] - [D_\nu, \Lam_\mu] + [B_{\mu\nu}, \lam]
- [F_{\mu\nu}, \del_5^{-1}\Lam_5^{\KK}],
\label{transf-Bij}
\eea
where the 5-dimensional covariant derivative and field strength are 
\bea
D_\mu &=& \del_\mu + A_\mu, \\
F_{\mu\nu} &=& [D_\mu, D_\nu],
\eea
with the gauge potential $A_\mu$ identified with 
the zero mode of $B_{\mu 5}$ through the relation
\be
A_\mu \equiv 2\pi R B_{\mu 5}^{\z}
= \oint dx^5 \; B_{\mu 5}.
\label{A-def}
\ee
The coefficient $2\pi R$ shows up from
the relation between the field theories on M5 and D4
and may be interpreted as the coupling constant $g$ on D4 \cite{r:HHM}.
The appearance of $\partial_5^{-1}$ in (\ref{transf-Bij}) 
is needed for a closed gauge symmetry algebra.

More explicitly, 
the transformation laws (\ref{transf-Bi5}), (\ref{transf-Bij})
can be decomposed into zero modes and KK modes as \cite{r:HHM}
\bea
\d A_{\mu} &=& [D_{\mu}, \lambda],
\\
\d B_{\mu 5}^{\KK} &=&
[D_\mu, \Lam_5^{\KK}] - \del_5 \Lam_\mu^{\KK} + [B_{\mu 5}^{\KK}, \lam],
\label{Bi5KK-transf}
\\
\d B_{\mu\nu}^{\z} &=&
[D_\mu, \Lam_\nu^{\z}] - [D_\nu, \Lam_\mu^{\z}] + [B_{\mu\nu}^{\z}, \lam],
\\
\d B_{\mu\nu}^{\KK} &=&
[D_\mu, \Lam_\nu^{\KK}] - [D_\nu, \Lam_\mu^{\KK}] + [B_{\mu\nu}^{\KK}, \lam]
- [F_{\mu\nu}, \del_5^{-1}\Lam_5^{\KK}].
\label{BijKK-transf}
\eea
This gauge symmetry algebra is closed.
It is \cite{r:HHM}
\be
[\d, \d'] = \d''
\ee
with the corresponding gauge parameters related via the following relations:
\bea
\lam'' &=& [\lam, \lam'], 
\\
{\Lam''_5}^{\KK} &=& [\lam, {\Lam'_5}^{\KK}] - [\lam', \Lam_5^{\KK}],
\\
\Lam''_{\mu} &=& [\lam, \Lam'_\mu] - [\lam', \Lam_\mu].
\eea

As the case of Abelian gauge symmetry for 2-form potentials,
there is a redundancy in using $\Lam_\mu$ and $\Lam_5$ 
to parametrize the non-Abelian gauge transformations defined above.
The gauge transformation is unchanged when $\Lam_\mu$ and $\Lam_5$
are changed by
\be
\d \Lam_\mu^{\KK} = [D_\mu, \xi^{\KK}], 
\qquad
\d \Lam_5^{\KK} = \del_5 \xi^{\KK}
\label{redundancy}
\ee
for an arbitrary function $\xi^{\KK}$ that has no zero mode.
Note that $\Lam_5^{\z}$ (equivalently $\lam$) is not transformed 
because it is the Wilson-loop degree of freedom 
of the gauge parameters.
This topological nature of $\lam$ is the qualification of 
its special role in the gauge transformation laws.

The field strength $H_{MNP}$ is defined by \cite{r:HHM}
\bea
H_{\mu\nu 5} &=& \frac{1}{2\pi R} F_{\mu\nu}
+ \del_5 B_{\mu\nu} 
+ [D_\mu, B_{\nu 5}^{\KK}] - [D_\nu, B_{\mu 5}^{\KK}], 
\label{Hmunu5-def} \\
H_{\mu\nu\kappa}^{\KK} &=& 
[D_\mu, B_{\nu\kappa}^{\KK}] + [D_\nu, B_{\kappa\mu}^{\KK}] 
+ [D_\kappa, B_{\mu\nu}^{\KK}]
\nn \\
&&
+ [F_{\mu\nu}, \del_5^{-1}B_{\kappa 5}^{\KK}] 
+ [F_{\nu\kappa}, \del_5^{-1}B_{\mu 5}^{\KK}] 
+ [F_{\kappa\mu}, \del_5^{-1}B_{\nu 5}^{\KK}].
\eea
In terms of the zero modes and KK modes,
eq. (\ref{Hmunu5-def}) is equivalent to
\bea
H_{\mu\nu 5}^{\z} &=& \frac{1}{2\pi R} F_{\mu\nu}, \\
H_{\mu\nu 5}^{\KK} &=& \del_5 B_{\mu\nu} 
+ [D_\mu, B_{\nu 5}^{\KK}] - [D_\nu, B_{\mu 5}^{\KK}].
\eea

All the components of the field strength $H$ defined above
transform covariantly in the form
\be
\d \Phi = [\Phi, \lam].
\label{cov-transf}
\ee
Although the definition of the component $H^{\z}_{\mu\nu\kappa}$ is missing,
luckily, in the self-dual gauge theory, 
we can completely ignore $H^{\z}_{\mu\nu\kappa}$ by focusing 
on its Hodge dual $H^{\z}_{\mu\nu 5}$ \cite{r:HHM}, 
which is essentially the ordinary Yang-Mills field strength 
$F_{\mu\nu}$ in 5 dimensions.
The self-duality condition for the zero modes is replaced 
by the Yang-Mills equation \cite{r:HHM}.

From the definitions of the field strengths, 
it is straightforward to derive the Bianchi identities \cite{r:HHM}:
\bea
{}\sum_{(3)} [D_{\mu}, H^{\z}_{\nu\kappa 5}]{} &=& 0, 
\\
{}\sum_{(3)} [D_{\mu}, H^{\KK}_{\nu\kappa 5}]{} &=& \del_5 H^{\KK}_{\mu\nu\kappa}, 
\\
{}\sum_{(4)} [D_{\mu}, H^{\KK}_{\nu\kappa\rho}]{} &=& 
\sum_{(4)} [F_{\mu\nu}, \del_5^{-1}H^{\KK}_{\kappa\rho 5}],
\eea
in which the 2-form potential $B_{MN}$ 
appears only through the field strength $H_{MNP}$
except the zero-mode $B^{\z}_{\mu 5}$ 
(or equivalently the 1-form potential $A_{\mu}$).
Here $\sum_{(3)}$ and $\sum_{(4)}$ refer to sums over 
permutations to totally anti-symmetrize all of the (3 or 4) indices in each expression.

The gauge symmetry defined above has the following properties:
\begin{enumerate}
\item
The gauge symmetry reduces to (multiple copies of) 
that for the Abelian 2-form gauge potential 
when the Lie algebra is Abelian.
\item
The ``gauge symmetry of gauge symmetry'' is consistently promoted to the non-Abelian case.
That is, the gauge transformation law (\ref{transf-Bi5}) and (\ref{transf-Bij}) 
parametrized by $\Lam_\mu, \Lam_5$ 
has the redundancy (\ref{redundancy}).
\end{enumerate}

It will be useful in computations below to define the covariant quantity
\be
\hat{B}_{\mu\nu} \equiv \del_5^{-1}H_{\mu\nu 5}^{\KK}.
\label{hat-B-def}
\ee
While $\hat{B}_{\mu\nu} = B_{\mu\nu}^{\KK}$ 
in the gauge $B_{\mu 5}^{\KK} = 0$,
the quantity $\hat{B}_{\mu\nu}$ transforms covariantly before gauge fixing.
In terms of $\hat{B}_{\mu\nu}$, 
the field strengths can be expressed as
\bea
H^{\KK}_{\mu\nu 5} &=& \del_5 \hat{B}_{\mu\nu}, \\
H^{\KK}_{\mu\nu\kappa} &=& 
[D_\mu, \hat{B}_{\nu\kappa}] + [D_\nu, \hat{B}_{\kappa\mu}] + 
[D_\kappa, \hat{B}_{\mu\nu}].
\eea

\subsubsection{Comment on Boundary Condition}
\label{comments}

Let us recall that, 
in gauge theories with one-form potentials,
Wilson loop arises as a new degree of freedom
when a spatial direction is compactified on a circle along $x^5$.
It can be represented by the zero mode of the gauge potential $A_5^{\z}$,
which behaves as a gauge-covariant scalar.
In our two-form gauge theory,
the analogue of $A^{\z}_5$ is $B_{\mu 5}^{\z}$, 
which behaves as a one-form potential in 5D.

Furthermore, 
when there is a compactification of an additional circle along $x^4$
in the one-form gauge theory,
it is possible to turn on a quantized flux on the torus of $(x^4, x^5)$.
It can be described by linear terms in the potential $A_4, A_5$.
Although linear terms are not periodic functions,
they are allowed because the potential only needs to be periodic 
up to gauge transformations.
Similarly, 
if we add a linear term to a periodic two-form potential $B_{MN}$ as
\bea
B_{\mu\nu} &\rightarrow& 
B'_{\mu\nu} = B_{\mu\nu} + \Sigma_{\mu\nu}(x) x^5,
\\
B_{\mu 5} &\rightarrow& B'_{\mu 5} = B_{\mu 5},
\eea
where $\Sigma_{\mu\nu}$ is independent of $x^5$.
If $B_{\m\n}$ is periodic,
$B'_{\m\n}$ is no longer of the form (\ref{decomp}) 
and satisfies the twisted boundary condition
\be
B'_{\m\n}(x^5+2\pi R) = B'_{\m\n}(x^5) + 2\pi R \, \Sigma_{\m\n}.
\ee
It is easy to check that
if the tensor $\Sigma_{\mu\nu}$ satisfies the relation
\be
\sum_{(3)} [D_{\mu}, \Sigma_{\nu\lam}] = 0,
\label{Bianchi}
\ee
all components of the new 3-form field strength,
\be
H'_{\mu\nu\lam} = H_{\mu\nu\lam},
\qquad
H'_{\mu\nu 5} = H_{\mu\nu 5} + \Sigma_{\mu\nu},
\ee
are still periodic as the old 3-form field strength.
In particular, all gauge-invariant quantities are periodic.
Incidentally,
the condition (\ref{Bianchi}) is equivalent to the Bianchi identity 
if $\Sigma_{\mu\nu}$ is proportional to $F_{\mu\nu}$.

By analogy,
the gauge transformation parameters $\Lam_M$, $\lam$
only need to be periodic up to the transformation (\ref{redundancy}).

Since the operator $\del_5^{-1}$ is well defined 
only when (\ref{decomp}) holds,
twisted boundary conditions requires an extension of 
our formulation on the 2-form potential.
We leave the complete theory including 
twisted boundary conditions for future works.

As a comment related to the issue of boundary conditions,
the non-Abelian self-dual gauge theory 
can be equivalently reformulated by 
adding a linear piece to $B_{\mu\nu}$ so that
\be
B_{\mu\nu} \equiv 
B^{\z}_{\mu\nu} + \frac{1}{2\pi R} F_{\mu\nu} x^5 + B^{\KK}_{\mu\nu}
\ee
and simplifying (\ref{Hmunu5-def}) by dropping the first term
\be
H_{\mu\nu 5} \equiv
\del_5 B_{\mu\nu} + [D_\mu, B_{\nu 5}^{\KK}] - [D_\nu, B_{\mu 5}^{\KK}],
\ee
while keeping all other definitions intact.
The linear term in $B_{\mu\nu}$ does not affect the periodicity of the field strength $H$.

\subsection{Lagrangian}
\label{Lagrangian}

The action for Abelian self-dual gauge fields
(also called ``chiral bosons'')
\cite{r:ChiralBosonLagrangian,r:PS,Chen:2010jgb,Pasti:2009xc,Huang:2011np}
can be found in various forms in the literature.
Having a non-Abelianized gauge symmetry for 2-form potentials,
one would like to construct a gauge-invariant action.

To write down a Lagrangian for self-dual gauge fields
in a manifestly Lorentz-covariant way,
one needs to introduce auxiliary fields. 
For simplicity, 
one often considers non-Lorentz-covariant expressions 
for Lagrangians without auxiliary fields.
They can be thought of as the gauge-fixed versions 
of certain Lorentz-covariant formulations.
For our purpose of describing M5-branes compactified on a circle,
the compactification partially breaks Lorentz covariance,
and it is natural to pick the compactified direction $x^5$
as a special direction in the Lagrangian formulation, 
with all Lorentz symmetry in the remaining directions $(x^0, x^1, \cdots, x^4)$ intact.

The action considered in Ref.\,\cite{r:HHM} for multiple M5-branes 
compactified on a circle 
is an extension of the Abelian version
\cite{Henneaux:1988gg}
\be
S =
- 
\int d^6 x \;
\tilde{H}^{\mu\nu 5}(H_{\mu\nu 5} - \tilde{H}_{\mu\nu 5})
\label{6D-action-1}
\ee
(up to an overall normalization),
where
\be
\tilde{H}_{\mu\nu5}
\equiv 
-\frac16 \epsilon_{\mu\nu\lambda\sigma\rho}H^{\lambda\sigma\rho}.
\ee

Decomposing the fields into zero modes and KK modes,
we note that this action for an Abelian 2-form potential 
is equivalent to
\be
S = 
-
\frac{1}{2\pi R} \int d^5 x \; 
\frac{1}{4} F_{\mu\nu} F^{\mu\nu}
-
\int d^6 x \;
\tilde{H}_{\KK}^{\mu\nu 5}(H^{\KK}_{\mu\nu 5} - \tilde{H}^{\KK}_{\mu\nu 5})
\ee
by suitably integrating out $H^{\z}_{\mu\nu\kappa}$
and redefining the gauge field $A_\mu$ via (\ref{A-def}).
The zero mode $H^{\z}_{\mu\nu\kappa}$ disappears from this action, 
but its existence is guaranteed by the Maxwell equation 
\footnote{
The Maxwell equation implies that there exits a tensor $B^{\z}_{\mu\nu}$ 
such that
\be
F_{\mu\nu} = \frac{1}{2}\eps_{\mu\nu\kappa\sigma\rho}
\del^{\kappa}B^{\z\sigma\rho}.
\label{F-sol}
\ee
One can then define $H^{\z}_{\mu\nu\kappa}$ by
\be
H^{\z}_{\mu\nu\kappa} = \del_{\mu}B^{\z}_{\nu\kappa}
+ \del_{\nu}B^{\z}_{\kappa\mu} + \del_{\kappa}B^{\z}_{\mu\nu},
\ee
and (\ref{F-sol}) is of the same form as the self-duality condition for the zero modes.
}
\be
\del^{\mu}F_{\mu\nu} = 0.
\ee
In this case,
the Maxwell equation is equivalent to the self-duality condition for the zero modes.

The action for the non-Abelian theory is then taken to be of the same form 
but with an overall trace \cite{r:HHM}
\be
S = 
-
\frac{1}{2\pi R} \int d^5 x \; 
\frac{1}{4} \Tr[F_{\mu\nu} F^{\mu\nu}]
- 
\int d^6 x \;
\Tr[\tilde{H}_{\KK}^{\mu\nu 5}(H^{\KK}_{\mu\nu 5} - \tilde{H}^{\KK}_{\mu\nu 5})],
\label{5D-action-1}
\ee
where all the fields are Lie-algebra valued.
As a generalization of the Abelian theory,
the Yang-Mills equation is by definition an equivalent expression 
of the self-duality condition on the zero modes.

To show that the self-duality condition for KK modes is equivalent to 
the equation of motion derived from this action, 
it is crucial to notice that,
in addition to the gauge symmetry for the 2-form potential 
(\ref{transf-Bi5}) and (\ref{transf-Bij}),
this theory has a new gauge symmetry
\be
\d B^{\KK}_{\mu\nu} = \Phi^{\KK}_{\mu\nu},
\qquad 
\d B_{\mu 5} = 0,
\label{add-gauge-1}
\ee
where $\Phi^{\KK}_{\mu\nu}$ satisfies the constraint 
\be
\eps^{\mu\nu\kappa\sigma\rho} [D_\kappa, \Phi^{\KK}_{\sigma\rho}] = 0.
\ee
This gauge symmetry is responsible for establishing 
the 1-1 correspondence between the equivalence classes of solutions 
to the equations of motion for the KK modes
\be
\epsilon^{\mu\nu\kappa\sigma\rho} 
[D_\kappa, (H^{\KK}_{\sigma\rho 5} - \tilde{H}^{\KK}_{\sigma\rho 5})] = 0
\ee
and the self-dual configurations defined by
\be
H^{\KK}_{\mu\nu 5} = \tilde{H}^{\KK}_{\mu\nu 5}.
\label{SD-KK}
\ee
Analogous additional gauge symmetries also appeared
in other M5-brane actions in the literature \cite{r:PS,r:PST}.
It is a universal feature of the Lagrangian formulation of chiral boson theories.

There are other equivalent formulations of Abelian self-dual gauge fields
that one can start with and extend it to the non-Abelian theory.
In particular, 
another choice of the action is
\be
S = 
-
\frac{1}{2\pi} \int d^6 x \;
H^{\mu\nu 5}(H_{\mu\nu 5} - \tilde{H}_{\mu\nu 5}),
\label{6D-action-2}
\ee
where $x^5 \in [0, 2\pi R)$,
as a small modification of the previous action (\ref{6D-action-1}).
It is different from (\ref{6D-action-1}) only in 
the first factor $H^{\mu\nu 5}$ of the Lagrangian.
We study this formulation in more detail now.

Like the previous action,
this action also enjoys an additional gauge symmetry
\be
\d B_{\mu\nu} = \Phi_{\mu\nu},
\label{add-gauge-2}
\ee
where $\Phi_{\mu\nu}$ is an arbitrary function independent of $x^5$.
This gauge symmetry implies that the zero mode of $B_{\mu\nu}$ 
is a pure gauge artifact.
The equation of motion for the KK modes derived from this new action is 
\be
\del_5\left(H^{\KK}_{\mu\nu 5} - \tilde{H}^{\KK}_{\mu\nu 5}\right) = 0,
\ee
and it is equivalent to the self-duality condition (\ref{SD-KK}).
The advantage of this choice is that 
the equivalence between equations of motion 
and self-duality condition is particularly simple.
It is also very easy to check that
the action (\ref{6D-action-2}) reduces directly to 
\be
S = 
-
\frac{1}{2\pi R} \int d^5 x \; 
\frac{1}{4} F_{\mu\nu} F^{\mu\nu}
- 
\int d^6 x \;
H_{\KK}^{\mu\nu 5}(H^{\KK}_{\mu\nu 5} - \tilde{H}^{\KK}_{\mu\nu 5}),
\label{5D-action-2-0}
\ee
without having to use the gauge symmetry (\ref{add-gauge-2})
or integrating out any field.

The non-Abelian counterpart of (\ref{5D-action-2-0}) is
\be
S = 
-
\frac{1}{2\pi R} \int d^5 x \; 
\frac{1}{4} \Tr[F_{\mu\nu} F^{\mu\nu}]
-
\int d^6 x \; 
\Tr[H_{\KK}^{\mu\nu 5}(H^{\KK}_{\mu\nu 5} - \tilde{H}^{\KK}_{\mu\nu 5})].
\label{5D-action-2}
\ee
Since the solutions to the equations of motion can be matched with 
self-dual configurations,
this Lagrangian is equivalent to the previous Lagrangian (\ref{5D-action-1})
at the classical level. 
It is not clear how they may be related to each other 
at the quantum level.
In general,
there are many classically equivalent Lagrangians 
for a self-dual gauge field \cite{Chen:2010jgb,Huang:2011np}.
It will be interesting to investigate the quantum theories for these actions.

\subsection{Canonical Formulation}

In this subsection,
we provide basics of the Lagrangian and Hamiltonian formulations
of the theory.

Let us repeat the Lagrangian (\ref{5D-action-2}) 
here for convenience of the reader:
\be
S =
- 
\frac{1}{2\pi R} \int d^5 x \; 
\frac{1}{4} \Tr[F_{\mu\nu} F^{\mu\nu}]
-
\int d^6 x \;
\Tr[H_{\KK}^{\mu\nu 5}(H^{\KK}_{\mu\nu 5} - \tilde{H}^{\KK}_{\mu\nu 5})].
\label{5D-action-2-1}
\ee
When the gauge-fixing condition
\be
B_{\mu 5}^{\KK} = 0
\label{gf}
\ee
is imposed,
this action is identical to the gauge field part of 
the supersymmetric action proposed in Ref.\,\cite{r:Bonetti}.

\subsubsection{Equation of Motion}

Note that $B_{\mu 5}^{\KK}$ appears in the action 
only through $\hat{B}_{\mu\nu}$ (\ref{hat-B-def}).
In terms of $A_\mu$ and $\hat{B}_{\mu\nu}$, 
the action (\ref{5D-action-2-1}) is
\be
S = 
-
\frac{1}{2\pi R} \int d^5 x \; 
\frac{1}{4} \mbox{Tr}[ F^{\mu\nu} F_{\mu\nu} ]
- 
\int d^6 x \; \mbox{Tr}[
\del_5\hat{B}^{\mu\nu} ( \del_5\hat{B}_{\mu\nu} 
+ \frac{1}{2}\eps_{\mu\nu\kappa\sigma\rho} [D^\kappa, \hat{B}{}^{\sigma\rho}] ) ].
\label{action}
\ee
The equation of motion for the KK modes $\hat{B}_{\mu\nu}$ is
\be
\del_5 \left( \del_5\hat{B}_{\mu\nu} 
+ \frac{1}{2}\eps_{\mu\nu\kappa\sigma\rho} [D^\kappa, \hat{B}^{\sigma\rho}] \right) = 0.
\ee
It is equivalent to
\be
\del_5\hat{B}_{\mu\nu} 
+ \frac{1}{2}\eps_{\mu\nu\kappa\sigma\rho} [D^\kappa, \hat{B}^{\sigma\rho}] = 0,
\label{SD}
\ee
as $\del_5^{-1}$ is well defined on KK modes.

The equation of motion for the zero modes $A_\mu$ is
\be
\frac{1}{2\pi R}[D_\nu, F^{\mu\nu}] + 
\frac{1}{2}\oint dx^5 \; \eps^{\mu\nu\kappa\sigma\rho} 
[\del_5\hat{B}_{\nu\kappa}, \hat{B}_{\sigma\rho}]
= 0.
\label{EOM-A}
\ee
This is of the form of the Yang-Mills equation with a source term.
It reduces to the pure Yang-Mills equation when KK modes vanish.

\subsubsection{Hamiltonian Formulation}

In the Lagrangian as well as the equations of motion,
the KK modes of $B_{\mu\nu}$ and $B_{\mu 5}$ 
are encoded in $\hat{B}_{\mu\nu}$,
and the zero modes are present in terms of $A_\mu$.
All physical gauge degrees of freedom in the theory 
reside completely in $\hat{B}_{\mu\nu}$ and $A_\mu$.

As there is no time-derivative terms of 
the temporal components $A_0$ and $\hat{B}_{0i}$ ($i, j = 1, 2, 3, 4$)
in the Lagrangian (\ref{5D-action-2-1}),
they are Lagrange multipliers.
The corresponding constraints are
\be
H^{\KK}_{0i5} = \tilde{H}^{\KK}_{0i5}
\label{constraint-0i}
\ee
for $\hat{B}^{\KK}_{0i}$ ($i, j = 1, 2, 3, 4$), 
and a modified Gauss' law for $A_0$.
As the canonical formulation of Yang-Mills theory is well known,
we will focus our attention on the KK modes.

The BRST anti-field formulation of the theory was already given in \cite{r:Kuo-Wei}.
Here we provide a simpler, more elementary Hamiltonian formulation.
To describe the Hamiltonian formulation for the KK modes,
we first solve the constraints (\ref{constraint-0i}),
which determine uniquely the values of the Lagrange multipliers
\be
\hat{B}_{0i} = - \frac{1}{6} \eps_{ijkl} \del_5^{-1} H^{jkl}_{(KK)}
\ee
in term of the dynamical fields $\hat{B}_{ij}$.
We can thus replace $\hat{B}_{0i}$ everywhere in the Lagrangian 
by this expression, 
so that the only dynamical fields of the KK modes are $\hat{B}_{ij}$.

As there is no more unsolved constraints,
we can define the conjugate momentum of $\hat{B}_{ij}$ simply as
\be
\hat{\Pi}_{ij} \equiv \frac{\d{\cal S}}{\d\del_0\hat{B}^{ij}}
= 
-
\frac{1}{2}\eps_{ijkl} H^{kl5}_{(KK)}.
\ee
Denoting the Fourier modes of a field $\Phi$
\be
\Phi = \sum_{n\in\mathbb{Z}} \Phi^{(n)} e^{-inx^5/R}
\ee
by $\Phi^{(n)}$ ($n\in\mathbb{Z}$),
the Poisson bracket is given by
\be
\{ \hat{B}_{ij}^{(m)}, \hat{B}_{kl}^{(n)} \} 
= i\frac{R}{n} \d^{m+n}_0 \epsilon_{ijkl}.
\ee
Here the superscripts $(m), (n)$ are labels 
for the KK modes.  

The Hamiltonian for the KK modes is
\be
{\cal H}^{\KK} = 
\int d^6 x \; \hat{\Pi}_{ij}\del_0\hat{B}^{ij} - S.
\ee
It can be simplified using self-duality conditions as
\be
{\cal H}^{\KK}
= - \int d^6 x \;
\left(
H_{0AB}^{\KK} H_{\KK}^{0AB}
\right) 
= 2 \int d^6 x \;
\left(
H_{0ij}^{\KK} H^{\KK}_{0ij}
+ H_{ijk}^{\KK} H_{ijk}^{\KK}
\right),
\ee
where $A, B = 1, 2, 3, 4, 5$ and $i, j = 1, 2, 3, 4$.
It is positive-definite.

\subsection{Conserved Currents}

Apart from the Hamiltonian, 
the momentum $P_5$ is also conserved due to 
translation symmetry in the $x^5$-direction.
The contribution of the KK-modes of the gauge field is  
\be
P_5^{\KK} =
\int d^5 x \;
\left(
H_{0ij}^{\KK} H_{\KK}^{ij5}
\right).
\ee
In fact, 
due to the property that KK modes only interact through zero modes,
there are infinitely many conserved charges.
For any positive integer $n$,
the KK modes labelled by $n$ and $-n$ can be simultaneously 
created or annihilated by a zero mode.
The number of excitations of the KK mode with label $n$ 
minus the number of excitations of the KK mode with label $-n$ is constant. 
There is thus a conserved current for each integer $n > 0$.

Formally, 
both actions (\ref{5D-action-1}) and (\ref{5D-action-2}) 
take the form $B^{(-n)} K B^{(n)}$
($K$ is an operator independent of  fields), 
so they are invariant under the transformation
\ba
\delta B^{(n)}=\epsilon_n B^{(n)},\qquad \delta B^{(-n)}=-\epsilon_{n} B^{(-n)}\,
\qquad (n > 0).
\ea
This is proportional to the transformation induced by a translation in $x^5$ 
if all parameters $\eps_n$ are given by $\eps_n = n\eps_1$.
But the transformation parameters $\eps_{n}$ $(n > 0)$
for different Fourier modes are allowed to be independent.
The translation symmetry in $x^5$ induces infinitely many symmetries 
because of the peculiar interaction feature of the theory.

These infinitely many symmetries lead to an infinite number of conserved currents,
\ba
j_{(n)}^\mu = \pi R \epsilon^{\mu\nu\lambda\rho\sigma} \mbox{Tr}\left(
H_{\nu\lambda5}^{(n)} B^{(-n)}_{\rho\sigma}-
H_{\nu\lambda5}^{(-n)} B^{(n)}_{\rho\sigma}
\right)=
n\pi i \mbox{Tr}\left(\epsilon^{\mu\nu\lambda\rho\sigma} 
B^{(n)}_{\nu\lambda} B^{(-n)}_{\rho\sigma}\right)\,
\ea
for $n=1,2,3,\cdots$.
The self-duality condition implies that they indeed satisfy
the conservation law $\partial_\mu j^\mu_{(n)} = 0$ in 5D.
$P_5$ is written in terms of them as (by taking $\epsilon_n=-in \epsilon/R$)
\be
P_5=-\sum_{n>0} \int d^5 x \; \frac{in}{R} \, j^0_{(n)}\,.
\ee

\section{Supersymmetry}
\label{SUSY}

A supersymmetric gauge theory in 5 dimensions 
for the gauge-fixed fields $A_{\mu}$ and $B^{\KK}_{\mu\nu}$ in the gauge (\ref{gf})
were proposed in Ref.\,\cite{r:Bonetti} to describe multiple M5-branes.
Like our formulation of the gauge theory for the 2-form potential, 
the zero modes and KK modes are treated separately in the supersymmetric theory.
We will show that 
the super-algbera in Ref.\,\cite{r:Bonetti}
can be viewed as the gauge-fixed version of
a super-algebra with the full gauge symmetry.
The extension of the supersymmetry to be fully consistent with the gauge symmetry
is necessary for the completeness of the M5-brane theory proposed in
Refs.\,\cite{r:HHM,r:Kuo-Wei,r:HoMatsuo}.


From the viewpoint of 5D SUSY, 
upon the compactification on a circle of radius $R$,
the massless fields on M5-branes is composed of 
the following SUSY multiplets \cite{r:Bonetti}:
\bea
(A^{\z}_\mu, \phi^{\z}, \chi^{\z}_a, Y^{\z}_{ab}) 
&=& \mbox{a massless vector multiplet}, \\
(F^{(n)}_{\mu\nu}, \phi^{(n)}, \chi^{(n)}_{a}, Y^{(n)}_{ab}) 
&=& \mbox{tensor multiplets with mass $m_n$}, \\
(h^{\z a\dot{b}}, \psi^{\z \dot{b}}) 
&=& \mbox{a massless hypermultiplet}, \\
(h^{(n) a\dot{b}}, \psi^{(n) \dot{b}}) 
&=& \mbox{hypermultiplets with mass $m_n$}.
\eea
The indices $a, b, \dot{a}, \dot{b}$
(taking values $1, 2$)
are the labels of the fundamental representations 
for two $SU(2)$ groups as part of the rotation symmetry 
of the transverse dimensions of the M5-branes.
The 5-dimensional uncompactified spacetime indices are $\mu, \nu = 0, 1, 2, 3, 4$.
The fermions $\chi^{(0)}_a, \psi^{(0)\dot{b}}$, $\chi^{(n)}_a, \psi^{(0)\dot{b}}$
are 5D spinors representing 6D Weyl spinors.

The mass of a field with KK-mode index $n$ is 
\be
m_n = \frac{n}{R},
\ee
and the auxiliary bosonic field $Y^{ab}$ has symmetrized indices: $Y^{ab} = Y^{ba}$.

In this theory of multiple M5-branes, 
all the fields are in the adjoint representation of the gauge group.
All the scalars $\phi, h_{a\dot{b}}$ and fermions $\chi_a, \psi_{\dot{b}}$
are covariant (\ref{cov-transf}) under gauge transformations.
The field $F^{(n)}_{\mu\nu}$ in Ref.\,\cite{r:Bonetti} should be identified 
with our gauge field strength through the relation
\be
F^{(n)}_{\mu\nu} \equiv R H^{(n)}_{\mu\nu 5}
= in \hat{B}^{(n)}_{\mu\nu}.
\label{FAB}
\ee 
In comparison with the notation of Ref.\,\cite{r:Bonetti},
other fields are also rescaled in a similar way.
\footnote{
In view of the 6D theory,
it is natural to rescale the fields in Ref.\,\cite{r:Bonetti},
which are labelled with the superscript $[BGH]$:
\be
\chi^{[BGH]}_{(n)a} = R \newchi_{(n)a},
\qquad
\phi^{[BGH]}_{(n)} = R \newphi_{(n)},
\qquad
Y^{[BGH]}_{(n)ab} = R \newY_{(n)ab}.
\ee
The variables on the right hand side
are those used in this paper.
}

We will use the totally antisymmetrized tensors
$\eps^{ab}, \eps^{\dot{a}\dot{b}}$ to raise or lower $SU(2)$ indices,
and we will use the NW-SE convention for contraction.
For example,
\be
\Phi^{a} = \eps^{ab} \Phi_{b}, 
\qquad
\Phi_{a} = \Phi^{b} \eps_{ba}.
\ee

There is an additional massless vector multiplet 
$(A^{\z}_{\mu}, \phi^{\z}, \chi^{\z}_a, Y^{\z}_{ab})$
defined in this model \cite{r:Bonetti}.
But it is fixed to be a constant
(see eq.(3.12) in Ref.\,\cite{r:Bonetti}),
and hence will be ignored.
Although only the 5D ${\cal N} = 2\;$ SUSY is manifest,
but we hope that the rest of the desired symmetry is hidden.
In fact, 
a method is proposed in Ref.\,\cite{r:Bonetti} to upgrade this model 
to another one with the full 6D ${\cal N} = (2, 0)$ superconformal symmetry.
We will focus on the simpler model in this work for clarity and simplicity.

The supersymmetry transformation laws
(eq. (4.22) in Ref.\,\cite{r:Bonetti}) are given by
\bea
\d A_\mu &=& - \frac{1}{2} \bar{\eps}^a \g_\mu \chi^{\z}_a, 
\label{SUSY-A} \\
\d \newphi^{(n)} &=& \frac{i}{2} \bar{\eps}^a \newchi^{(n)}_{a}, 
\label{SUSY-phi} \\
\d H^{(n)}_{\mu\nu 5} &=& \bar{\eps}^a \g_{[\mu} D_{\nu]}\newchi^{(n)}_{a}
- \frac{i}{2} [\newphi^{(n)}, \bar{\eps}^a \g_{\mu\nu} \chi^{\z}_a]
+ \frac{i}{2} \bar{\eps}^a \g_{\mu\nu} (D_{\phi} \newchi^{(n)}_{a}), 
\label{SUSY-F} \\
\d \newchi^{(n)a} &=& \frac{1}{4} \g^{\mu\nu} H^{(n)}_{\mu\nu 5} \eps^a 
- \frac{i}{2} \Ds \newphi^{(n)} \eps^a - \newY^{(n)ab}\eps_b
- \frac{1}{2} (D_{\phi} \newphi^{(n)}) \eps^a, 
\label{SUSY-chi} \\
\d \newY^{(n)ab} &=& - \frac{1}{2} \bar{\eps}^{(a} \Ds \newchi^{(n)b)}
+ i [\newphi^{(n)}, \bar{\eps}^{(a}\chi^{b)}] 
- \frac{i}{2} \bar{\eps}^{(a} (D_{\phi} \newchi^{(n)b)}), 
\label{SUSY-Y} \\
\d h^{(n)a\dot{b}} &=& - i \bar{\eps}^a \psi^{(n)\dot{b}}, 
\label{SUSY-h} \\
\d \psi^{(n)\dot{b}} &=& \frac{i}{2} \Ds h^{(n)a\dot{b}} \eps_a
+ \frac{1}{2} (D_{\phi} h^{(n)a\dot{b}}) \eps_a,
\label{SUSY-psi}
\eea
where we have used the notation $D_{\phi}$ defined by
\be
(D_{\phi} \Phi^{(n)}) \equiv - i m_n \Phi^{(n)} + [\phi^{(0)}, \Phi^{(n)}],
\label{D-phi}
\ee
and 
$\Phi_{[\mu\nu]} \equiv \frac{1}{2}(\Phi_{\mu\nu} - \Phi_{\nu\mu})$,
$\Phi^{(ab)} \equiv \frac{1}{2}(\Phi^{ab} + \Phi^{ba})$
for symmetrized and anti-symmetrized indices.
The covariant derivative is 
$D_\mu = \del_\mu + A_\mu$.
\footnote{
The convention in Ref.\,\cite{r:Bonetti} is that 
$D_\mu = \del_\mu - A_\mu$.
As a result, $A_\mu, F_{\mu\nu}$ here differ from those in Ref.\,\cite{r:Bonetti}
by a sign.
}
For any field $\Phi$, 
its zero mode is denoted as $\Phi^{\z}$.
All the equations above are valid for 
$n = 0$ (the zero modes) as well.

Notice that the zero mode of the scalar $\phi^{(0)}$ appears 
only through the operator $D_{\phi}$  
in the gauge transformation laws.
(The same is true for the Lagrangian.)
It is tempting to interpret 
$D_{\phi}$ as the covariant derivative $D_5$ 
in the Fourier basis, 
and $\phi^{(0)}$ as the (missing) 5th component $A_5$ 
of the 1-form gauge potential.
It is peculiar that a transverse coordinate $\phi^{(0)}$ 
of the M5-brane also resembles a component of the 1-form potential 
upon compactification. 

Our task here is to find the SUSY transformation law for
the component $B_{MN}^{\KK}$,
which is absent in the (gauge-fixed) SUSY transformation laws 
(\ref{SUSY-A})--(\ref{SUSY-psi}).
The SUSY transformation of $A^{(0)}_{\mu}$ (\ref{SUSY-A})
more or less suggests that,
before gauge fixing,
\bea
\d B^{(n)}_{\mu 5} &=& - \frac{1}{2} \bar{\eps}^a \g_\mu \newchi^{(n)}_{a}.
\eea
In addition, 
the SUSY transformation law (\ref{SUSY-F})
for the gauge-covariant field $H_{\mu\nu 5}^{(n)}$
suggests that we define the gauge transformation 
of the rest of the gauge potential components $B_{\mu\nu}^{(n)}$ by
\bea
\d B^{(n)}_{\mu\nu} &=& 
- \frac{i}{2} \bar{\eps}^a \g_{\mu\nu} \newchi^{(n)}_{a}
- \frac{R}{2n} [\newphi^{(n)}, \bar{\eps}^a \g_{\mu\nu} \chi^{\z}_a]
+ \frac{R}{2n} [\phi^{\z}, \bar{\eps}^a \g_{\mu\nu} \newchi^{(n)}_{a}]
- \frac{iR}{n} [B^{(n)}_{[\mu 5}, \bar{\eps}^a \g_{\nu]} \chi^{\z}_a].
\nn \\
\eea

To summarize,
the SUSY transformation laws for the zero modes 
are the same as that for the 5D super Yang-Mills theory,
and the SUSY transformation laws for the KK modes are given by
\bea
\d B^{\KK}_{\mu 5} &=& 
- \frac{1}{2} \bar{\eps}^a \g_\mu \newchi^{\KK}_{a},
\nn \\
\d B^{\KK}_{\mu\nu} &=& 
- \frac{i}{2} \bar{\eps}^a \g_{\mu\nu} \newchi^{\KK}_{a}
- \frac{i}{2}[\del_5^{-1} \newphi^{\KK}, \bar{\eps}^a \g_{\mu\nu} \chi^{\z}_a]
+ \frac{i}{2}[\phi^{\z}, \bar{\eps}^a \g_{\mu\nu} \del_5^{-1}\newchi^{\KK}_{a}]
+ [\del_5^{-1} B^{\KK}_{[\mu 5}, \bar{\eps}^a \g_{\nu]} \chi^{\z}_a],
\nn \\
\eea
together with (\ref{SUSY-phi}) and (\ref{SUSY-chi})
--(\ref{SUSY-psi}).

Let us check whether the super-algebra 
for the SUSY transformations defined above 
is closed up to gauge transformations.
It is straightforward to check that SUSY transformations
on $B^{(n)}_{\mu 5}$ satisfy the closure relation
\be
[\d_1, \d_2] B^{(n)}_{\mu 5} = 
\a^\nu \del_\nu B^{(n)}_{\mu 5} 
+ \b \frac{in}{R} B^{(n)}_{\mu 5} + [D_\mu, \Lam^{(n)}_5]
+ [B^{(n)}_{\mu 5}, \lam] - \frac{in}{R}\Lam^{(n)}_{\mu},
\label{ddA}
\ee
or equivalently,
\be
[\d_1, \d_2] B^{\KK}_{\mu 5} = 
\a^\nu \del_\nu B^{\KK}_{\mu 5} + \b \del_5 B^{\KK}_{\mu 5} 
+ [D_\mu, \Lam^{\KK}_5] 
- \del_5 \Lam^{\KK}_{\mu}
+ [B^{\KK}_{\mu 5}, \lam] ,
\label{ddA-1}
\ee
where the coefficients are given by
\bea
\a^\mu &=& \frac{1}{2} \bar{\eps}_2^a \g^\mu \eps_{1a}, 
\label{alpha} \\
\b &=& \frac{i}{2} \bar{\eps}_2^a \eps_{1a},
\label{beta} \\
\Lam^{\KK}_5 &=& - \a^\mu B^{\KK}_{\mu 5} 
+ \frac{\b}{R} \phi^{\KK}, 
\label{Lambda_5} \\
\Lam^{\KK}_{\mu} &=& 
\a^\nu B^{\KK}_{\mu\nu} + \b B^{\KK}_{\mu 5} 
+ \frac{\a_\mu}{R} \phi^{\KK}
+ \left[\del_5^{-1}\left(\b B^{\KK}_{\mu 5} + \frac{\a_\mu}{R} \phi^{\KK}\right), 
\phi^{\z}\right],
\nn \\
&& 
\label{Lambda_i} \\
\lam &=& - R \a^\mu B^{\z}_{\mu 5} + \b \phi^{\z}.
\label{lambda}
\eea
On the right hand side of (\ref{ddA-1}),
the first term is a translation in the $x^\mu$ direction, 
the second term is a translation in the $x^5$ direction,
the third and fourth terms are gauge transformations by $\Lam^{\KK}_i, \Lam^{\KK}_5$
and the last term is a gauge transformation by $\lam$
(the 5D gauge transformation parameter).
The gauge transformation pieces in the super-algebra agree nicely with 
the gauge transformation of $B_{\mu 5}^{\KK}$ (\ref{Bi5KK-transf}).

It can be checked that the same super-algebra 
observed for $B_{\mu 5}^{\KK}$ in (\ref{ddA-1}), that is,
\be
[\d_1, \d_2] = 
\a^i p_i + \b p_5 
+ \d_{\Lam} + \d_{\lam}
\label{super-algebra}
\ee
applies to all other fields, 
with the parameters $\a^i$, $\b$, $\Lambda$ and $\lambda$
defined by (\ref{alpha}) -- (\ref{lambda}).
Here $(\d_1, \d_2)$ are the SUSY transformations with parameters $(\eps_1, \eps_2)$,
$\d_{\Lam}$ is the gauge transformation for the KK modes of the 2-form potential,
$\d_{\lambda}$ is the 5D SYM gauge transformation,
and $p_i$, $p_5$ are generators of translations, 
which are for our case just derivatives $\del_i$, $\del_5$.

\section{Solitonic Solutions}
\label{Soliton}

All BPS states invariant under translation along $x^5$
survives dimensional reduction and 
can be represented by configurations in the 5D SYM theory.
They are all automatically included in the theory studied here,
including those discussed in Refs. \cite{r:LPS} and \cite{Tachikawa:2011ch}.
In the following, 
we will look for BPS solutions involving KK modes.

According to the SUSY transformation laws (\ref{SUSY-A}) -- (\ref{SUSY-psi}),
a BPS configuration for the KK modes should allow nontrivial solutions of 
the SUSY parameter $\epsilon$ to the following equations
\bea
0 &=& \frac{1}{4} \g^{\mu\nu} H^{\KK}_{\mu\nu 5} \eps^a 
- \frac{i}{2} \Ds \phi^{\KK} \eps^a - Y_{\KK}^{ab}\eps_b
- \frac{1}{2} D_{\phi}\phi^{\KK} \eps^a,
\label{BPS-1}
\\
0 &=& \frac{i}{2} \Ds h_{\KK}^{a\dot{b}} \eps_a
+ \frac{1}{2} D_{\phi}h_{\KK}^{a\dot{b}} \eps_a,
\label{BPS-2}
\eea
assuming that all fermionic fields vanish.
Here the derivative $D_{\phi}$ (\ref{D-phi})
and its complex conjugate are defined by
\bea
(D_{\phi} \Phi^{(n)}) \equiv - i m_n \Phi^{(n)} + [\phi^{(0)}, \Phi^{(n)}],
\\
(\bar{D}_{\phi} \Phi^{(n)}) \equiv + i m_n \Phi^{(n)} + [\phi^{(0)}, \Phi^{(n)}].
\label{Db-phi}
\eea


In various circumstances,
\footnote{
For example, 
see \cite{Ho:2012dn}.
}
the BPS conditions are not sufficient to guarantee 
the satisfaction of all equations of motion.
Hence we list here for reference the equations of motion for the KK modes 
derived from the supersymmetric action of Ref.\,\cite{r:Bonetti}:
\begin{align}
&\frac{n}{R}H^{\m\n 5}_{(n)} - \frac{i}{2} \eps^{\m\n\lam\s\r} D_{\lam} H_{(n)\s\r 5}
- i ([\phi_{(0)}, H_{(n)}^{\m\n 5}] - [\phi_{(n)}, F^{\m\n}])
= 0, 
\label{EoM-H}
\\
&D_{\m} D^{\m} \phi_{(n)} - \frac{n^2}{R^2} \phi_{(n)}
+ \frac{iR}{2n}[F^{\m\n}, H_{(n)\m\n 5}]
- \frac{iR}{n}[D^{\m}\phi_{(0)}, D_{\m}\phi_{(n)}]
- \frac{iR}{n}[\phi_{(0)}, D^{\m} D_{\m}\phi_{(n)}] &
\nn \\
&\qquad 
- \frac{2iR}{n}[Y^{ab}_{(0)}, Y_{(n)ab}]
- \frac{in}{R}[\phi_{(0)}, \phi_{(n)}]
- [\phi_{(0)}, [\phi_{(0)}, \phi_{(n)}]]
- \frac{iR}{n} [\phi_{(0)}, [\phi_{(0)}, [\phi_{(0)}, \phi_{(n)}]]]
= 0, 
\label{EoM-phi}
\\
&Y_{(n)ab} - \frac{iR}{n}([\phi_{(0)}, Y_{(n)ab}]-[\phi_{(n)}, Y_{(0)ab}])
= 0, 
\label{EoM-Y}
\\
&D^{\m} D_{\m} h_{(n)a\dot{b}} - \frac{n^2}{R^2} h_{(n)a\dot{b}}
- [h^b_{(n)\dot{b}}, Y_{(0)ab}] - \frac{2in}{R}[\phi_{(0)}, h_{(n)a\dot{b}}]
+ [\phi_{(0)}, [\phi_{(0)}, h_{(n)a\dot{b}}]] = 0. 
\label{EoM-h}
\end{align}
In the above we have set all fermions to zero.

In terms of $D_{\phi}$ (\ref{D-phi}) and $\bar{D}_{\phi}$ (\ref{Db-phi}), 
they are simplified as
\begin{align}
&\bar{D}_{\phi}H_{(n)}^{\m\n 5} - \frac{i}{2} \eps^{\m\n\lam\s\r} D_{\lam} H_{(n)\s\r 5} 
+ i[\phi_{(n)}, F^{\m\n}] = 0, 
\\
&\bar{D}_{\phi}(D^{\m} D_{\m} \phi_{(n)} + D_{\phi}D_{\phi} \phi_{(n)})
+ \frac{iR}{2n}[F^{\m\n}, H_{(n)\m\n 5}]
- \frac{iR}{n}[D^{\m}\phi_{(0)}, D_{\m}\phi_{(n)}]
- \frac{2iR}{n}[Y^{ab}_{(0)}, Y_{(n)ab}] = 0, 
\\
&\bar{D}_{\phi}Y_{(n)ab} - [\phi_{(n)}, Y_{(0)ab}] = 0,
\\
&(D^{\m} D_{\m} + D_{\phi}D_{\phi})h_{(n)a\dot{b}}
- [h^b_{(n)\dot{b}}, Y_{(0)ab}] = 0.
\end{align}

\subsection{M2 Along $x^4$}
\label{M2alongx4}

An M2-brane stretched between two M5-branes separated by a finite distance 
in a transverse direction
intersects with either M5-brane on a one-dimensional space,
and it is described as a self-dual string
from the viewpoint of the M5-brane worldvolume theory.
The description for these states is known for the zero modes (in SYM theory) \cite{r:LPS},
however this description may not be complete.
If the self-dual string lies along the $x^5$-direction, 
it can certainly be described by zero modes.
But if it lies along other directions, say $x^4$,
the zero modes can only describe the state 
when the self-dual string is smeared over 
the circle along $x^5$.
We will consider the extension of these zero-mode BPS solutions 
by turning on KK modes, 
in order to describe a self-dual string that 
is localized in the $x^5$-direction.
Our strategy is to first find zero-mode BPS solutions, 
and then consider small fluctuations of the KK modes 
with the zero-mode solution as a background,
ignoring back-reactions.

\subsubsection{Zero-Mode Solution}

If an M2-brane is not wrapped around $x^5$,
but lies along $x^4$,
it is described by a static string-like configuration
\cite{r:LPS}
\be
F_{i'j'} = \eps_{i'j'k'} D^{k'} \Phi, 
\qquad
A_0 = 0,
\qquad
A_4 = \sin \th \; \Phi, 
\qquad
X^6 = \cos \th \; \Phi,
\label{solution-x4}
\ee
where $i', j', k' = 1, 2, 3$ and
$\Phi$ satisfies $D^2 \Phi = 0$.
To regulate the total energy and momentum, 
we compactify $x^4$ on a circle of radius $R_4$.
Let $r \equiv \sqrt{\sum_{i'=1}^3 x_{i'}^2}$.
At large $r$, the solution of $\Phi$ is approximately
\be
\Phi = \phi_0 \s^3 - \frac{q\s^3}{4\pi r} + \cdots,
\label{phi3}
\ee
where $\phi_0$ is an arbitrary constant 
and $q \in \mathbb{Z}$.
For
\bea
\cos\th = \frac{v/2}{\sqrt{v^2/4 + 4\pi^4 n^2/q^2}}, 
\qquad
\sin\th = \frac{2\pi^2 n/q}{\sqrt{v^2/4 + 4\pi^4 n^2/q^2}},
\qquad
\phi_0 = \sqrt{v^2/4 + 4\pi^4 n^2/q^2}, 
\nn \\
\eea
the momentum and magnetic charge are
\be
P_5 = - 2\pi R_4 \frac{4\pi^2 n}{g_{YM}^2}, 
\qquad
Q_{M4} = \frac{vq}{g_{YM}^2},
\ee
where
the YM coupling is related to the radius of the circle of $x^5$ by
\be
\frac{4\pi^2}{g_{YM}^2} = \frac{1}{R}.
\ee
The energy is
\be
E = 2\pi R_4 \sqrt{Q_{M4}^2 + (P_5/2\pi R_4)^2}.
\ee
This solitonic solution preserves half of the SUSY 
in the 5D SYM theory.

%
%

\remove{
Let us extend the solution above to include KK modes.
Consider the ansatz:
\be
H^{\KK}_{ij5} = f(x^0, x^5) F_{ij}, 
\qquad
\phi^{\KK} = f(x^0, x^5) \cos\th \Phi, 
\qquad
Y^{\KK}_{ab} = h^{\KK}_{a\dot{b}} = 0
\label{ansatz-2}
\ee
for the same gauge potential $A_{i}$ and function $\Phi$ in the above solution.
The function $f(x^0, x^5)$ is arbitrary except that 
it integrates to zero over $x^5$.
}

In Ref.\,\cite{r:LPS},
it is claimed that 
the zero-mode solution above
represents all the BPS configurations for 
a self-dual string winding around the circle of $x^4$.
This has to be the case if the 5D SYM theory is indeed 
the complete description of multiple M5-branes.
In our approach,
while there are independent KK-mode degrees of freedom,
one may wonder if it is possible to find BPS states 
in which KK modes are excited on top of this zero-mode configuration
so that the self-dual string is not uniformly smeared over $x^5$.

\subsubsection{KK-Mode Solution}
\label{M2-x4-KK}

Despite the lack of a complete theory with Lagrangian and SUSY transformation rules,
field equations for the 2-form gauge potential and 
BPS conditions were proposed in Ref.\,\cite{Chu:2012rk} 
for M2-branes ending on multiple M5-branes.
A solution exists \cite{Chu:2012rk} to 
represent an open M2-brane stretched between two M5-branes, 
lying along the $x^4$-direction, 
with an $x^5$-dependent distribution.
(It can be extended to more general solutions for more than two M5-branes \cite{Chu:2013hja}.)
The similarity and differences between the theory of Ref.\,\cite{Chu:2012rk}
with our theory of multiple M5-branes will be discussed later in Sec. \ref{compare-2},
but coincidentally the solution found in Ref.\,\cite{Chu:2012rk}
can be adopted for our calculation.
(We will see later that the equations considered in Ref.\,\cite{Chu:2012rk}
are only a subset of all the equations one needs to check
in the model studied here.)

For simplicity, 
we consider the special case of $\th = 0$ 
in (\ref{solution-x4}) for the zero modes
\be
F_{i' j'} = -i \eps_{i' j' k'}D^{k'} \phi^{\z},
\qquad
A_0 = A_4 = 0, 
\label{ansatz-M2-x4-0-mode}
\ee
where $i', j', k' = 1, 2, 3$.
This implies that $F_{0 i'} = F_{04} = F_{i' 4} = 0$.
For the purpose of including KK modes
in a way that will be convenient for our discussions below,
let us re-calculate the zero-mode solution 
by starting with 
the ansatz for the 't Hooft-Polyakov monopole:
\bea
A_{i'} &=& \eps_{i' j' k'} f(r) x^{j'} \sigma_{k'}, 
\label{ansatz-A-x4} \\
\phi^{\z} &=& h(r) x\cdot \sigma,
\label{ansatz-phi-x4}
\eea
where $x\cdot\sigma \equiv x^{i'}\sigma_{i'}$,
\be
r \equiv \sqrt{x^{i'}x^{i'}}
\ee
and $\sigma_{i'}$ represents generators of the $su(2)$ Lie algebra
with the commutation relations
\be
{}[\s_{i'}, \s_{j'}] = \eps_{i' j' k'}\s_{k'}.
\ee
(Repeated indices are summed over even when 
they are both subscripts or both superscripts.)

Eq. (\ref{ansatz-M2-x4-0-mode}) then implies that
\bea
&\frac{1}{r}\frac{df}{dr} + f^2 = \frac{1}{r}\frac{dh}{dr} + fh, 
\label{diff-eq-1}
\\
&\frac{1}{r}\frac{df}{dr} + \frac{2}{r^2}f = fh 
- \frac{1}{r^2}h.
\label{diff-eq-2}
\eea
(These two equations can be combined to give 
a single (non-linear) second order differential equation for $h(r)$.)
An explicit solution to these equations was given in Ref.\,\cite{Chu:2012rk}:
\bea
f(r) &=& 
\frac{1}{r^2} - \frac{c}{r\sinh(cr)},
\\
h(r) &=& 
\frac{1}{r^2} - \frac{c}{r} \coth(cr)
\label{sol-h}
\eea
with a constant parameter $c$.
The solution above is singular at $r = 0$, 
the location of the M2-brane.
The fact that it has to be singular somewhere 
is expected from the equation 
\be
D_{i'}D^{i'}\phi^{\z} = 0,
\ee
which can be derived by taking covariant derivative on 
the first equation of (\ref{ansatz-M2-x4-0-mode}),
since the second order differential operator $D^2$ is negative definite.

To compare this solution with the expression (\ref{phi3}) in the previous subsection,
one can carry out a gauge transformation 
\be
\phi \rightarrow U \phi U^{-1},
\qquad
F_{i' j'} \rightarrow U F_{i' j' } U^{-1}
\ee
by the $SU(2)$ matrix
\be
U \equiv \exp\left[-\frac{(x^1 \s_2 - x^2 \s_1)}{\sqrt{(x^1)^2 + (x^2)^2}}
\tan^{-1}\left(\frac{\sqrt{(x^1)^2 + (x^2)^2}}{x^3}\right)\right],
\label{U}
\ee
which is also singular at the origin to bring it 
to the form in which 
$\phi \rightarrow (c - \frac{1}{r}) \s_3$
at large $r$. 

In fact, 
we will not need the explicit solution for the discussion below.
All we need is that the zero-mode solution 
can be put in the form (\ref{ansatz-A-x4}) and (\ref{ansatz-phi-x4}).

For the KK modes, as above, 
we focus on solutions with $h_{a\dot{b}}^{(n)} = 0$.
First, we assume that all interaction terms vanish 
in the equations of motion 
(to be verified later)
by demanding
\begin{align}
&[\phi_{(0)}, H_{(n)}^{\m\n 5}] - [\phi_{(n)}, F^{\m\n}] = 0, 
\label{assume-1}
\\
&[F^{\m\n}, H_{(n)\m\n 5}]
- 2[D^{\m}\phi_{(0)}, D_{\m}\phi_{(n)}]
= 0,
\\
&[\phi_{(0)}, \phi_{(n)}] = 0, 
\\
&Y_{(n)ab} = 0, 
\label{assume-2}
\end{align}
so that the equations of motion are linearized
\begin{align}
&\frac{n}{R}H^{\m\n 5}_{(n)} - \frac{i}{2} \eps^{\m\n\lam\s\r} D_{\lam} H_{(n)\s\r 5}
= 0, 
\label{EoM-H-2}
\\
&D_{\m} D^{\m} \phi_{(n)} - \frac{n^2}{R^2} \phi_{(n)} = 0.
\label{EoM-phi-2}
\end{align}

We also extend the BPS conditions (\ref{ansatz-M2-x4-0-mode})
for the zero-mode solution to the KK modes by 
\be
H^{(n)}_{i' j' 5} = -i \eps_{i' j' k'}D^{k'} \phi^{(n)},
\qquad
\hat{B}^{(n)}_{0i'} = \hat{B}^{(n)}_{i'4} = 0.
\label{ansatz-M2-x4-KK-mode}
\ee
According to the BPS conditions for the KK modes (\ref{BPS-1}), (\ref{BPS-2}),
this ansatz (\ref{ansatz-M2-x4-KK-mode}) preserves
1/4 of the SUSY 
for SUSY parameters $\eps^a$ satisfying the conditions
\be
\g_5 \eps^a = - \eps^a, 
\qquad
\g_{04} \eps^a = \eps^a.
\ee
(Recall that the solutions for M2-branes wrapped around $x^5$
are also 1/4-BPS states.)

Eq. (\ref{ansatz-M2-x4-KK-mode}) implies that
$H^{(n)}_{0 i' 5} = H^{(n)}_{i' 45} = 0$.
($H^{(n)}_{045}$ will not be zero.)
The self-duality condition then implies that
$H^{(n)}_{0 i' j'} = H^{(n)}_{i' j' 4} = 0$.

The equations of motion (\ref{EoM-H-2}) and (\ref{EoM-phi-2})
would be valid if
\begin{align}
&\hat{B}^{(n)}_{04} = \phi^{(n)}, 
\label{B04=phi} \\
&D_{i'}D^{i'}\phi^{(n)} = \frac{n^2}{R^2}\phi^{(n)}.
\label{DDphi=phi}
\end{align}
What we need to do now is to find explicit solutions 
for (\ref{DDphi=phi}).
Then we can determine the values of 
$\hat{B}_{i' j'}^{(n)}$ and $\hat{B}_{04}^{(n)}$
using (\ref{ansatz-M2-x4-KK-mode}) and (\ref{B04=phi}).

Following (\ref{ansatz-A-x4}) and (\ref{ansatz-phi-x4}),
we take the ansatz
\be
\phi^{(n)} = h^{(n)}(r) x\cdot \sigma
\label{ansatz-phi-n-x4}
\ee
to find solutions to the equation (\ref{DDphi=phi}),
which implies that $h^{(n)}$ satisfies the equation
\be
\frac{d^2 h^{(n)}}{dr^2} 
+ \frac{4}{r}\frac{dh^{(n)}}{dr} + 4f h^{(n)} - 2 r^2 f^2 h^{(n)}
= \frac{n^2}{R^2}h^{(n)}.
\label{diff-eq-3}
\ee
An explicit solution to this equation was found in Ref.\,\cite{Chu:2012rk}:
\be
h^{(n)}(r) = c_n\frac{e^{-|n|r/R}}{r^2}\left(
1 + \frac{cR}{|n|} \coth(cr)
\right)
\label{sol-hn}
\ee
for arbitrary parameters $c_n$.
Since all KK modes are decoupled from all other KK modes,
we have infinitely many parameters $c_n$ to
parametrize the amplitude of each KK mode independently.

It can now be checked that
the assumptions (\ref{assume-1})--(\ref{assume-2}) are valid.
As only the ansatz
(\ref{ansatz-A-x4}), (\ref{ansatz-phi-x4}) and (\ref{ansatz-phi-n-x4})
are needed for this check,
all solutions of $f(r), h(r), h^{(n)}(r)$ to the differential equations
(\ref{diff-eq-1}), (\ref{diff-eq-2}) and (\ref{diff-eq-3}) 
give legitimate BPS states in the multiple M5-brane theory.

Note that the zero-mode solution in SYM theory
discussed in the previous subsection also represents an M2-brane along $x^4$,
but it is smeared over the circle of $x^5$.
Here we have found the solutions 
with an arbitrary distribution along $x^5$, 
including those localized around a point in the $x^5$-direction.
This allows us to consider the localization of the M2-brane 
in all transverse directions.

\subsubsection{Infinite $R$ Limit}
\label{Infinite-R}

We take the BPS solution above for M2-branes in the $x^4$-direction
as an example to demonstrate how the theory of multiple M5-branes 
for finite radius $R$ can also be used to obtain information about infinite $R$,
the uncompactified space.
%

In the limit of small $R$,
the zero mode $\phi^{(0)}$ dominates over the KK modes.
For a localized source in three large transverse dimensions $(x^1, x^2, x^3)$,
the massless field $\phi^{(0)}$ should scale as $1/r$
with $r = \sqrt{(x^1)^2+(x^2)^2+(x^3)^2}$ at small $r$, 
when the kinetic term dominates over the potential term in its field equation.
This is indeed the case in the solution of $\phi^{\z}$ above 
in (\ref{ansatz-phi-x4}) and (\ref{sol-h}).
(Note that 
the factor $(x\cdot \s)/r$ can be gauge-transformed to $\s_3$ via $U$ (\ref{U}).)
Similarly,
the KK modes $\phi^{(n)}$ behave as massive fields in three large transverse dimensions
and scales like $e^{-|n|r/R}/r$ in the UV limit.
This can be verified by examining the solution
of $\phi^{(n)}$ in (\ref{ansatz-phi-n-x4}) and (\ref{sol-hn}).
%

On the other hand, 
for a large radius $R$ of the compactified circle,
$\phi$ should behave as a massless field in four large transverse dimensions $(x^1, x^2, x^3, x^5)$.
Hence one expects that,
in the UV limit when the field equation is dominated by the kinetic term,
$\phi$ scales like $1/\rho^2$ ($\rho = \sqrt{r^2 + (x^5)^2}$) 
as a result of rotation symmetry in $(x^1, x^2, x^3, x^5)$.
Note that, 
since the 5D Lorentz symmetry in $(x^0, x^1, x^2, x^3, x^4)$
is manifest in the theory,
this rotation symmetry implies the full 6D Lorentz symmetry.
%

In the limit of large $R$, 
it is more convenient to replace the index $n$ for KK modes 
by the wave number 
\be
k \equiv \frac{n}{R}.
\label{k}
\ee
In this limit, 
the sum over KK modes $\sum_n$ is approximated by
an integral over $k$:
\be
\sum_{n\in\mathbb{Z}} F(n) \simeq R \int_{-\infty}^{\infty} dk F(Rk)
\ee
for any function $F(n)$.
For a delta-function source at $x^5 = 0$,
we superpose all $KK$ modes with equal amplitude 
since $\int dk \; e^{-ikx^5} = 2\pi \delta(x^5)$.
That is, 
we choose $c_n = \a$ to be independent of $n$ 
in the solution for each KK mode (\ref{sol-hn}),
and sum over $n$ to find
\bea
\phi 
&=& 
\sum_{n\in\mathbb{Z}} \phi^{(n)} e^{-in x^5/R}
\nn \\
&\simeq&
\a R \int dk \; \frac{1}{r} e^{-|k|r-ikx^5} 
\left(1+
\frac{c}{|k|}
\coth(cr)
\right) \frac{x\cdot\s}{r}
\nn \\
&=&
\frac{2\a R}{\rho^2}\frac{x\cdot\s}{r}
-\frac{2\a R c}{r} \coth(cr) \log(\rho/\Lambda) \frac{x\cdot\s}{r},
\eea
where $\Lambda$ is an IR cut-off parameter, 
and the factor
$\frac{x\cdot\s}{r}$ can be transformed to $\s^3$
by a gauge transformation through the matrix (\ref{U}).
We should take $\a \rightarrow 0$ as $R\rightarrow \infty$
such that the solution $\phi$ is finite in the limit of large $R$.
%

In the UV limit,
$\phi$ is dominated by the first term which indeed 
demonstrates the $1/\rho^2$ behavior implied by the 6D Lorentz symmetry.
The second term in the expression of $\phi$
depends on the parameter $c$ which characterizes the profile 
of the soliton solution in the $(x^1, x^2, x^3)$-directions.
Since we have taken a Dirac $\delta$-function profile for the solution 
in the $x^5$-direction,
we do not expect this term to be invariant under the 4D rotations
in $(x^1, x^2, x^3, x^5)$.
For a nontrivial evidence of the 6D Lorentz symmetry,
one should find a solution (with a nontrivial $x^5$-profile)
invariant under the 4D rotation
at finite $r$ in the large $R$ limit.

\subsection{BPS States for Pure KK Modes}

Since all KK modes interact only with zero modes, 
they are all decoupled if we set all zero modes to zero.
The system becomes equivalent to an infinite set of free fields.

Setting the zero modes to zeros,
the equations of motion (\ref{EoM-H})--(\ref{EoM-h})
are simplified to
\ba
i*_5 dF_{(n)} - m_n F_{(n)} = 0, \label{EoM-0H} \\
(\del^{\m}\del_{\m}+m_n^2)\phi_{(n)} = 0, \\
(\del^{\m}\del_{\m}+m_n^2)h_{(n)a\dot{b}} = 0, \\
Y_{(n)ab} = 0,
\ea
where $F_{(n)} \equiv \frac{1}{2} RH_{(n)\m\n 5} dx^{\m} \wedge dx^{\n}$
is a two-form in 5D,
and $*_5$ denotes the Hodge dual in 5D.

Even though all the KK modes are decoupled in the equations of motion,
they are related by the BPS conditions for a BPS state.
The BPS conditions (\ref{BPS-1}) and (\ref{BPS-2}) are simplified as
\bea
&-\frac14 H_{(n)\mu\nu 5}\gamma^{\mu\nu}\epsilon_a
-\frac{i}2 \gamma^\mu\partial_\mu \phi_{(n)}\epsilon_a
-Y^{ab}_{(n)}\epsilon_b+\frac{im_n}{2}\phi_{(n)}\epsilon_a = 0,
\label{BPS-01}
\\
&\g^{\mu}\del_{\mu} h_{\KK}^{a\dot{b}} \eps_a
- \frac{n}{R} h_{\KK}^{a\dot{b}} \eps_a = 0.
\label{BPS-02}
\eea
In general it relates the KK modes $H_{(n)\mu\nu 5}, \phi_{(n)}$
and $Y_{(n)ab}$ to one another.

\subsubsection{M-Waves}
\label{wave}

There are KK modes representing uniform sinusoidal waves 
propagating along the $x^5$ direction are BPS states. 
These M-waves solutions that we will present below 
were first obtained \cite{Chu:2013joa}
for a different proposal of the M5-brane theory \cite{r:ChuKo}.
But here these solutions are to be checked against 
the field equations and BPS conditions 
of a complete theory with a supersymmetric Lagrangian 
and gauge symmetry.

Consider the ansatz of self-dual configurations
\be
H^{ij5}_{(n)} = \frac{1}{2} \eps^{ijkl} H_{(n)kl5}.
\ee
All equations of motion are satisfied by
\be
H_{(n)ij5} = c_{(n)ij} e^{in x^0/R},
\qquad
\phi_{(n)} = \mbox{const}\times e^{in x^0/R},
\qquad
h_{(n) a\dot{b}} = \mbox{const}\times e^{in x^0/R},
\ee
where $c_{(n)ij}$ is a self-dual constant matrix
\be
c_{(n)ij} = \frac{1}{2}\eps_{ijkl}\, c_{(n)kl},
\ee
and the equation of motion (\ref{EoM-0H})
implies that $H_{(n) 0i5} = 0$.
There are no relations among the amplitudes 
as all KK modes are decoupled.

These independent waves of gauge fields and scalars 
are 1/4-BPS states symmetric for SUSY parameters $\eps^a$ satisfying
\be
\g_{1234} \eps^a = \eps^a, 
\qquad
\g^{0} \eps^a = -i \eps^a.
\ee
Obviously these solutions also survive in the large $R$ limit 
by replacing the KK mode index $n$ by $k$ (\ref{k}).

\section{Supersymmetric Gerbe}
\label{s:gerbe}

The discussion in the previous sections can be straightforwardly
generalized to the set-up in Ref.\,\cite{r:HoMatsuo} where 
a formulation of non-Abelian gerbes was proposed.
Let $G$ be an arbitrary Lie group and $\rho$ be
an arbitrary (not necessarily irreducible) representation.
We write $\ttg$ to represent the Lie algebra of $G$ and 
$\ur$ to be the representation of $\ttg$.
Let $V$ be the representation space where $\ur$ acts.  
$V$ can be regarded as an Abelian group by the action of addition. 
For the example of $N$ M5-branes,
$G = U(N)$ and
$V$ is the space of KK modes in the adjoint representation.

In this set-up, we define the one-form $(A,\tA)$ to take values
in the semi direct product $\ttg \ltimes V$ $(A\in \ttg, \ \tA\in V)$
and the two-form $B\in V$. The pair $\ttg \ltimes V$ 
and $V$ is an example of  crossed module,
which is the standard ingredient to define a non-Ablian gerbe. 
In Ref.\,\cite{r:HoMatsuo}, we argued that a system
with the structure of non-Abelian gerbe is often limited to free or
topological theory. 
Indeed, 
such topological theory was used to classify 
the phases of non-Abelian gauge theory in four dimensions 
\cite{Gukov,Kapustin}.
Our example seems to be the only exception
where some modification of the gauge transformation enables
us to define an interacting field theory.

We also need to include a mass matrix $M$ which is a linear map
acting on $V$ and commute with the action of $\ur$.  
Suppose $V$ is decomposed into the invariant subspaces 
$V=\oplus_i V_i$,
so that $M=\oplus_i m_i I_i$,
where $I_i$ is an identity matrix acting on $V_i$.
Our discussion so far corresponds to a specific choice
$V=\oplus_{n=1}^\infty (V_n\oplus V_{-n})$, 
$m_n=n/R$ and $V_{\pm n}$ is the adjoint representation of $\ttg$.

\subsection{Gauge Transformation}

We introduce the zero-form gauge parameters $\Lambda\in \ttg$ and $\tL\in V$
and the one-form gauge parameter $\ta\in V$.
The gauge transformation proposed in Ref.\,\cite{r:HoMatsuo} is,
\ba
\delta A_i&=& \partial_i \Lambda+[A_i,\Lambda],
\label{ddA1}\\
\delta \tilde A_i &=& \partial_i\tilde\Lambda +\ur(A_i)(\tilde\Lambda)
-\ur(\Lambda)(\tA_i)+M\ta_i,
\label{ddA2}\\
\delta \tB_{ij}&=& \partial_i\ta_j-\partial_j\ta_i+\ur(A_i)(\ta_j)-\ur(A_j)(\ta_i)
-\ur(\Lambda)(\tB_{ij})+M^{-1}\ur(F_{ij})(\tL)\,.
\label{ddB}
\ea
The last term in (\ref{ddB}) is a modification necessary to have 
homogeneous gauge transformation of field strength,
\ba
F_{ij}&=& \partial_i A_j-\partial_j A_i+[A_i, A_j],
\\
\tF_{ij}&=& \partial_i \tA_j-\partial_j\tA_i+\ur(A_i)(\tA_j)-\ur(A_j)(\tA_i)-M\tB_{ij},
\label{ddtF}\\
\tZ_{ijk}&=& \sum_{(3)}\left(\partial_i \tB_{jk}+\ur(A_i)\tB_{jk}-M^{-1} \ur(F_{ij})(\tA_k)\right),
\label{defZ2}
\ea
such that $\bF_{ij}=(F_{ij}, \tF_{ij})\in \ttg\ltimes V$.

Transformation of curvature becomes
\ba
\delta F_{ij}&=& [F_{ij},\Lambda]\,,\\
\delta \tF_{ij}&=& -\ur(\Lambda)(\tF_{ij})\,,\\
\delta \tZ_{ijk}&=& -\ur(\Lambda)(\tZ_{ijk})\,.
\ea

In order to see the correspondence with the
previous sections, one may consider taking
one of $V_n$ in 
$V=\oplus_{n=1}^\infty (V_n\oplus V_{-n})$
and translate the notation
in the previous sections by the following rules:
\footnote{
Note that $A_{\mu}$ and $F_{\mu\nu}$ in this section are different from 
$A_{\mu}$ and $F_{\mu\nu}$ in other sections of this paper by a factor of $2\pi R$.
}
\ba
&&B_{\mu5}^{(0)}\rightarrow A_\mu,\quad
B_{\mu5}^{(KK)}\rightarrow \tA_\mu,\quad
B_{\mu\nu}^{(KK)}\rightarrow i\tB_{\mu\nu}\\
&& \Lambda_5^{(0)}\rightarrow \Lambda,\quad
\Lambda_5^{(KK)}\rightarrow \tilde\Lambda,\quad
\Lambda_\mu^{(KK)}\rightarrow i\ta_\mu\,,\\
&& H_{\mu\nu5}^{(KK)}\rightarrow \tF_{\mu\nu},\quad
H_{\mu\nu\kappa}^{(KK)}\rightarrow i\tilde Z_{\mu\nu\kappa}\,,
\ea
with $\partial_5 \rightarrow i M$ ($M=m_n$).
We use different indices $i,j$ instead of $\mu,\nu$ since
parts of this section can be applicable to other dimensions.

\subsection{Action for non-Abelian Gerbe}

The homogeneity of the gauge transformations
enables us to write the gauge invariant action,
\ba\label{LFtFtZ}
L=-\frac14 \mbox{Tr}(F_{ij})^2-\frac14 \langle\tF_{ij},\tF^{ij}\rangle
-\frac{1}{12}\langle\tZ_{ijk},\tZ^{ijk}\rangle\,.
\ea
Here $\langle\cdot,\cdot\rangle$ is an inner product
in $V$ which is invariant under the action of $G$.

For our interest in self-dual gauge theories,
a covariant action which leads to the self-dual equation is
\ba\label{action2}
S=\int d^5 x \; \mbox{Tr} \; \tF^{\mu\nu}(\tF_{\mu\nu}-i(*\tZ)_{\mu\nu}).
\ea
The gauge field part of our action for the multiple M5-branes 
is a special case of this expression.
The equation of motion derived from (\ref{action2}) is
\ba\label{sd1}
\tF-i*\tZ=0.
\ea

\subsection{SUSY Partners and Transformation Laws in 5 dimensions}

SUSY relates $A_i$ with $\chi_a, \phi$ and $\tB, \tA$ with $\tchi,\tphi$,
so $\chi_a, \phi\in g$ and $\tchi,\tphi\in V$.
In Ref.\,\cite{r:Bonetti}, 
SUSY transformation closes without the extra gauge parameters $\tL, \ta$,
so it is natural to define the fermion transformation
to be homogeneous as the field strength,
\ba
\delta \chi_a = [\chi_a,\Lambda],\qquad
\delta \tchi_a = -\ur(\Lambda)\tchi_a\,.
\ea
The gauge transformations of $Y,\phi,\tphi,\tilde h,\tilde\psi$ should be similar to $\chi,\tchi$.


The SUSY transformation laws for the general case of non-Abelian gerbes
are a straightforward extension of the SUSY transformation laws 
(\ref{SUSY-A})--(\ref{SUSY-psi}) given the special case of multiple M5-branes
first given in Ref.\,\cite{r:Bonetti}.
They are
\ba
\delta A_\mu &=& -\frac12 \bar\epsilon^a \gamma_\mu \chi_a,
\\
\delta \tA_\mu &=& -\frac12 \bar\epsilon^a\gamma_\mu \tchi_a,
\\
\delta \tB_{\mu\nu} &=& -\frac{1}2 \bar\epsilon^a\gamma_{\mu\nu}\tchi_a
- \frac{i}{2M} \bar\epsilon^a\gamma_{\mu\nu}\left(\ur(\chi_a)(\tphi)
+ \ur(\phi)\tchi_a\right)
+\frac{1}{M} \bar\epsilon^a\gamma_{[\nu}\ur(\tchi_a)\tA_{\mu]},
\\
\delta\chi_a&=& \frac14 \gamma^{\mu\nu} F_{\mu\nu}\epsilon^a-
\frac{i}2 \sD\, \phi \epsilon^a - Y^{ab}\epsilon_b -\frac12 D_\phi \phi \epsilon^a,
\\
\delta \tchi^a &=& \frac14 \gamma^{\mu\nu} \tF_{\mu\nu}\epsilon^a-
\frac{i}2 \sD\, \tphi \epsilon^a - \tY^{ab}\epsilon_b -\frac12 D_\phi \tphi \epsilon^a,
\\
\delta \tY_{ab}&=& -\frac12 \bar\epsilon^{(a} \sD \tchi^{b)}+i\ur(\tphi)(\bar\epsilon^{(a}\chi^{b)})
-\frac{i}{2} (D_\phi\bar\epsilon^{(a}\tchi^{b)}),
\ea
where $D_\phi \tilde\Phi=-iM \tilde\Phi +\ur(\phi)\tilde \Phi$.
Thus we see that the 5D supersymmetric gauge theory for multiple M5-branes 
allows us to choose any non-Abelian gerbe defined in \cite{r:HoMatsuo}.

\section{Comments}
\label{Comment}

\subsection{On KK Modes}

Some \cite{r:Douglas,r:LPS} proposed that 
the same 5-dimensional SYM theory for D4-branes 
can be interpreted as a theory for M5-branes 
even at finite radius.
It was claimed that
all momentum modes on M5-branes are described by
zero-mode configurations with non-zero 4D instanton charges.
This proposal attracted a lot of attention and was investigated by many 
(see for example
\cite{Singh:2011id,r:Kim,Linander:2011jy,Kim:2012ava,Kim:2012tr,Bern:2012di}).
On the other hand, 
we believe that,
although the 4D instantons carry $P_5$-charge, 
there are other independent KK-mode degrees of freedom.
The KK modes should be kept explicitly in the M5-brane theory.
Our arguments are as follows.

Roughly speaking, 
for a matter field $\Phi$, 
the momentum density $p_5$ is of the form
\be
\Pi \del_5 \Phi,
\label{p5-Phi}
\ee
where $\Pi$ is the conjugate momentum of $\Phi$.
In the free field theory of a single M5-brane,
the momentum density $p_{5}$ due to $B_{ij}$ is
proportional to
\be
H_{0ij}H^{ij5} = \frac{1}{2}\eps_{ijkl}H^{kl5}H^{ij5}.
\label{p5-B}
\ee
While the zero mode contribution 
\be
\frac{1}{2}\eps_{ijkl}H^{kl5}_{\z}H^{ij5}_{\z} = \frac{1}{2}\eps_{ijkl}F^{ij}F^{kl}
\ee
of the 2-form potential $B_{ij}$
is indeed the instanton density, 
the question is whether 
the KK mode contribution in (\ref{p5-Phi}) and (\ref{p5-B})
should all be discarded in the multiple M5 theory.
If we accept the single M5-brane theory as a correct low energy effective theory
(which can be verified by studying solitonic solutions corresponding to M5-branes
in the 11 dimensional supergravity),
both instantons on D4-branes and KK-modes of matter fields
contribute to the momentum $p_5$ through (\ref{p5-Phi}) and (\ref{p5-B}), 
at least in the limit when all M5-branes are far apart and decoupled.

Some may argue that 
KK modes in M theory are identified with D0-branes in string theory, 
and D0-branes are identified with instantons on D4-branes, 
so KK modes are equivalent to instantons.
However,
the identification of D0-branes with instantons 
is justified only in the low energy, small $R$ limit,
because the SYM theory is only a low energy effective theory 
in the limit of small $R$.
More precisely, 
D4-brane is the KK reduction of M5-brane compactified on a small circle.
As KK reduction removes KK modes, 
zero-modes carrying $P_5$ charge
(such as instantons)
survive KK reduction and persist in the D4-brane theory.
The fact that D0-branes on D4-branes can be identified with instantons
does not imply that all D0-branes are described as instantons
before taking the low energy, small $R$ limit.

To claim that the instanton configurations of gauge fields accounts for 
all possible sources of $P_5$ requires a new type of gauge symmetry in which 
the KK-mode degrees of freedom is gauge-equivalent to the instanton configurations.
There has never been such an example in field theory.

Finally, 
without the KK modes,
it would be hard to imagine how one can describe 
the BPS states we considered in Sec. \ref{M2-x4-KK},
that is, 
parallel M2-branes lying in the $x^4$-direction
(or other large spatial directions)
when they are not uniformly smeared over in the $x^5$-direction.

\subsection{On Zero-Modes}
\label{compare-2}

A central idea in our formulation is to identify the vector field $A_{\mu}$
needed in a non-Abelian gauge theory 
with certain components of the tensor field $B_{MN}$
by choosing a special direction 
(the compactified direction $x^5$),
to avoid excessive physical degrees of freedom.
After the proposal of Ref.\,\cite{r:HHM},
a similar strategy was taken in Ref.\,\cite{r:ChuKo},
followed by a series of publications \cite{Chu:2012rk,Chu:2013hja,Chu:2013joa,Chu:2013gra}.
We explain here the differences between their model and ours,
and hopefully through this discussion 
the reader will also understand our model better,
in particular about the zero mode sector.

The main difference of Ref.\,\cite{r:ChuKo} from our proposal 
lies in the treatment of the zero modes of $B_{MN}$, 
which leads to a difference in the equation of motion for $A_{\mu}$.
In our approach, 
the equation of motion for $A_{\mu}$ reduces to the standard 5D YM equations 
when we set all KK modes to zeros.
This is not the case for the theory proposed in Ref.\,\cite{r:ChuKo}.

In the discussion below,
we label a quantity defined in Ref.\,\cite{r:ChuKo} by the symbol ``[CK]''.
The work of Ref.\,\cite{r:ChuKo} defined the 1-form potential $A_{\m}^{[CK]}$ via the equation
(Eq.(3.19) in Ref.\,\cite{r:ChuKo}):
\be
F_{\mu\nu}\CK = \int dx^5 \tilde{H}\CK_{\mu\nu 5},
\label{FH}
\ee
where $\tilde{H}\CK_{\mu\nu 5}$ is defined as 
(denoted as $\tilde{H}_{\mu\nu}$ in eq.(3.2) of Ref.\,\cite{r:ChuKo})
\be
\tilde{H}\CK_{\mu\nu 5} = 
\frac{1}{2}\eps_{\mu\nu\kappa\sigma\rho}[D^{\kappa}, B^{\sigma\rho}\lCK].
\ee
In Ref.\,\cite{r:ChuKo},
eq.(\ref{FH}) restricts the zero-mode $B^{\z\CK}_{\mu\nu}$.
In contrast,
the zero-modes in our model are defined only in terms of $B^{\z}_{\mu 5}$, 
without explicitly referring to $B^{\z}_{\mu\nu}$.

The problem with eq.(\ref{FH}),
or the reason why we have avoided explicit reference 
to $B^{\z}_{\m\n}$ in our model,
is its deviation from the Yang-Mills equation
when KK modes are removed on dimensional reduction.
Taking the covariant derivatives on both sides of (\ref{FH}),
we get
\be
[D^{\nu}, F_{\mu\nu}\CK] = \frac{1}{4} \int dx^5 
\eps_{\mu\nu\kappa\sigma\rho}[F^{\nu\kappa}\lCK, B^{\sigma\rho}\lCK].
\label{DF=FB-1}
\ee
After removing all KK modes,
the Yang-Mills equation is still modified by a term of the form 
$\frac{\pi R}{2} [F\lCK, B^{\z}\lCK]$
on the right hand side.
(Note that $B^{\z}\lCK$ is constrained by (\ref{FH})
so one cannot set it to zero at will.)
They need to prove that 
somehow the correction term is negligible 
in the low energy limit in order for
their model to be consistent with D4-brane physics.

Another difference is that, 
in our model,
we have a free 1-form parameter $\Lam$ for 
the non-Abelian gauge transformations,
while it is strongly constrained to 
a much smaller gauge symmetry in Ref.\cite{r:ChuKo}.
In fact, 
if one does not demand the explicit presence of such a gauge symmetry 
and an invariant action at the same time, 
the no-go theorem \cite{r:Bekaert1,r:Bekaert2,r:Bekaert3} would not be applicable, 
and the introduction of nonlocality may not be fully justified.

Incidentally, 
despite their claim,
the 6D Lorentz symmetry in the model of Ref.\,\cite{r:ChuKo} 
is not a genuine Lorentz symmetry in the usual sense,
as the definition of the angular momentum
involves an integral over the whole space-time.
Furthermore,
their proposed Lorentz transformation can be defined even 
after adding more symmetry-breaking terms in the Lagrangian. 

An interesting question is whether it is possible 
to write down an uncompactified theory for multiple M5-branes.
There are strong constraints 
on the $S$-matrix \cite{r:MM5b} for 
self-interactions of the self-dual tensor multiplet
from Lorentz symmetry and supersymmetry in 6D.
On the other hand,
various physical aspects of the uncompactified theory 
can be extracted in the large $R$ limit of the compactified theory,
as we did in Sec. \ref{Infinite-R}.
An uncompactified theory is not in crucial need unless
it has some advantages such as manifest covariance 
in Lorentz symmetry, supersymmetry and gauge symmetry. 

\subsection{Conclusion}

In addition to the works mentioned above,
there are many other attempts 
to formulate an effective theory for multiple M5-branes,
or just to explore potentially interesting higher-form gauge theories in 6D.
Some approached the problem through the mathematical notion of 3-algebra
\cite{r:LP,Kawamoto:2011ab,Papageorgakis:2011xg,Lambert:2011gb,Lambert:2012qy},
higher gauge theory or twistor space
\cite{r:Aschieri,r:SaemannWolf,Saemann:2013pca,Palmer:2013pka,Palmer:2013haa}.
Some used holographical principle as a tool
\cite{Fiorenza:2012tb,Minahan:2013jwa,Mori,Palmer:2014jma}.
The interest in multiple M5-brane theory has also inspired 
new theoretical frameworks for higher gauge theories
\cite{r:SSW,Samtleben:2012fb,Bandos:2013jva,Bandos:2013sia},
which are interesting by themselves.

The model studied in this paper based on 
\cite{r:HHM,r:Kuo-Wei,r:HoMatsuo,r:Bonetti} satisfies
the following criteria for an effective theory of multiple M5-branes:
(i) It agrees with 5D SYM in the absence of KK modes.
(ii) It agrees with 6D single M5-brane when the gauge group is Abelian.
(iii) It has the full gauge symmetry for a 2-form potential.
(iv) It has the correct field content.
It is the only model satisfying all of those requirements.
However,
only part of the supersymmetry,
and part of the rotation symmetry in the transverse directions 
of the M5-brane are manifest.
The full 6D Lorentz symmetry in the UV limit 
is also not yet proven.
%

More tests on the model should be carried out,
especially on its hidden Lorentz symmetry and supersymmetry.
It will also be interesting to study the large $R$ limit in more detail,
including scattering processes,
and to compare the results with the no-go conclusion
based on supersymmetry of Ref.\,\cite{r:MM5b}.

We believe that a good comprehension of the multiple M5-brane system 
will be considered a significant breakthrough not only in string theory,
but also in the context of general field theories, 
as it will open a door to a new class of symmetries 
and related new physics 
that we know very little of.

\section*{Acknowledgement}

The authors would like to thank 
Heng-Yu Chen, Chong-Sun Chu, Kazuo Hosomichi, Yu-Tin Huang, Takeo Inami
for their interest and discussions.
PMH is supported in part by
the National Science Council, Taiwan, R.O.C.
and by the National Taiwan University
NSC-CDP-102R3203.
YM is partially supported by Grant-in-Aid
(KAKENHI \#25400246)
from MEXT Japan.


\end{CJK} 

\end{document}